\documentclass[a4paper,12pt]{article}

\usepackage[utf8]{inputenc}
\usepackage[T1]{fontenc}
\usepackage{authblk}
\usepackage[dvips]{graphicx}
\usepackage{amsmath,amssymb}
\usepackage{mathptmx}
\usepackage{booktabs}
\usepackage{bm}
\usepackage[dvips]{color}
\usepackage{float}
\usepackage[dvipdfmx]{hyperref}
\usepackage{hyperref}
\usepackage{color}

\title{Monopoles, spectra of overlap fermions, and eta-prime meson in external magnetic fields}

\author{M. Hasegawa\thanks{E-mail: hasegawa@theor.jinr.ru}}

\affil{Bogoliubov Laboratory of Theoretical Physics, Joint Institute for Nuclear Research, Dubna, Moscow Region 141980, Russia}

\date{\today}

\begin{document}

\maketitle
\begin{abstract}
  The effects of external magnetic fields on monopoles, spectra of the overlap Dirac operator, instantons, and the mass of the eta-prime meson are examined by conducting lattice QCD simulations. The uniform external magnetic fields are applied to gauge field configurations with $N_{f}$ = 2 + 1 flavor quarks. The bare quark masses are tuned in order to obtain the physical values of the pion mass and of the ratio $\frac{m_{s}}{m_{u, d}}$. Standard configurations and configurations with applied external magnetic fields are generated in the color confinement and deconfinement phases. The intensity of external magnetic fields varies from $e|B|$ = 0.57 to 1.14 [GeV$^{2}$]. To examine the influence of external magnetic fields on monopoles, we first calculate the monopole density, measure the lengths of the monopole loops and compare them with the absolute value of the Polyakov loops. Next, using the generated configurations, we compute the eigenvalues and eigenvectors of the overlap Dirac operator, which preserves exact chiral symmetry. To investigate how external magnetic fields affect the spectra of the overlap Dirac operator, we compute spectral densities, compare fluctuations of the eigenvalues of the overlap Dirac operator with the predictions of random matrix theory, and estimate the number of instantons and anti-instantons from the topological charges. In addition, we analyze smearing effects on these observables and chiral symmetry breaking. Finally, we calculate the decay constant of the pseudoscalar meson, the chiral condensate, and the square mass of the eta-prime meson using the eigenvalues and eigenvectors. We then extrapolate the numerical results in the chiral limit and demonstrate the effects of external magnetic fields on the extrapolation results. This article presents preliminary results.
\end{abstract}

%%%%%%%%%%%%%%%%%%%%%%%%%%%%%%%%%%%%%%%%%%%%%%%%%%%%%%%%%%%%%%%%%%%%%%%%%%%
%% SEC 1
%%%%%%%%%%%%%%%%%%%%%%%%%%%%%%%%%%%%%%%%%%%%%%%%%%%%%%%%%%%%%%%%%%%%%%%%%%%

\section{Introduction}

Magnetic monopoles play a critical role in the color confinement mechanism explained by the phenomenological model proposed by 't Hooft and Mandelstam~\cite{tHooft2,Mandelstam1}. Phenomenological models of instanton liquid~\cite{Shuryak2} and instanton vacuum~\cite{Dyakonov6} demonstrate how instantons lead to chiral symmetry breaking. Magnetic monopoles and instantons in the QCD vacuum are closely intertwined, connecting color confinement and chiral symmetry breaking~\cite{Hasegawa2,Hasegawa4}. Furthermore, the eta-prime meson has a close association with topological charges, instantons, and monopoles, as noted in references~\cite{Giusti8,DeGrand3,Hasegawa3}.

In the early universe, cosmological phase transitions led to the creation of a primordial magnetic field that affects all gauge fields corresponding to the unbroken symmetry group~\cite{Vachaspati_1}. Therefore, magnetic fields are assumed to impact quantum chromodynamics (QCD).

The chiral magnetic effect~\cite{Kharzeev_1} refers to the occurrence of electric charge separation along the direction of the magnetic field due to chirality imbalance. In QCD matter, the imbalance in chirality originates from transitions between distinct topological states. The imbalance manifests itself as alterations in topological charges, as determined by the difference in chirality of fermion zero modes. It causes non-dissipative transport of electric charges between varying chirality states.

Thus, it is important to examine the chiral magnetic effect using overlap fermions which preserve exact chiral symmetry in lattice gauge theory.

Recently, the construction of the Nuclotron-based Ion Collider fAcility (NICA) in Dubna in Russia has been underway for studying Quark-Gluon Plasma (QGP), the phase transition of color confinement and deconfinement, as well as the phase transition of the breaking and restoration of chiral symmetry~\cite{NICA}. It is essential to analyze the impact of strong magnetic fields that arise during the collision of heavy-ion beams in experiments on observables.

The effects of magnetic fields on strongly interacting matter were investigated through the studies of lattice QCD by introducing external magnetic fields into the gauge field configurations. A notable example is the investigations of {\it magnetic catalysis}, which demonstrates that the magnetic fields enhance chiral symmetry breaking~\cite{Bali1,Bali2,Massimo2,Massimo3}.

The inhabitants of the QCD vacuum, monopoles and instantons, are closely related to color confinement and chiral symmetry breaking~\cite{Hasegawa2,Hasegawa4}. Therefore, it is interesting to study the effects of external magnetic fields on topology of the QCD vacuum. However, due to the strong interaction at low-energy of QCD, it is difficult to demonstrate the effects on topology of the QCD vacuum by perturbative calculations. Furthermore, it is widely acknowledged that conducting lattice QCD simulations near the chiral limit and utilizing the fermion action that maintains the exact chiral symmetry are exceedingly time-intensive endeavors. Consequently, the fermion action, which does not conserve chiral symmetry, is primarily employed to generate gauge configurations set to heavy quark masses.

Therefore, the Pisa group generates gauge field configurations with $N_{f}$ = 2 + 1 flavor quarks in the framework of the Pisa-Dubna collaboration\footnote{The configurations were generated by Francesco Negro in the Pisa group.}. The bare quark masses are tuned in order to obtain the physical values of the pion mass and of the ratio $\frac{m_{s}}{m_{u, d}}$. The standard configurations, as well as the configurations to which the uniform external magnetic fields are applied, are generated at low (50 [MeV]) temperature in the color confinement phase and high (200 [MeV]) temperature in the color deconfinement phase~\cite{Massimo2,Massimo3}. The intensity of the uniform external magnetic fields ranges from $e|B|$ = 0.57 to 1.14 [GeV$^{2}$]. 

Moreover, we calculate the eigenvalues and eigenvectors of the massless overlap Dirac operator that holds exact chiral symmetry in lattice gauge theory~\cite{Ginsparg1, Neuberger1, Neuberger2, Lusher1, Chandrasekharan1} using these configurations. We use approximately 400 eigenvalues and eigenvectors for each conﬁguration, which is approximately four times the amount utilized in our previous study~\cite{Hasegawa2,Hasegawa4}. As a result, the numerical outcomes include higher energy level contributions.

First, we measure the lengths of the monopole loops and create histograms to represent their distribution. The length of the longest monopole loops correlates with the temperature at which color deconfinement occurs~\cite{Hasegawa4}, and also with the number of instantons and anti-instantons~\cite{DiGH4}. To show the effects of external magnetic fields on the monopoles in the color confinement and deconfinement phases, we calculate the mean absolute values of the Polyakov loops and compare them with the lengths of the longest monopole loops.

Second, we investigate the impacts of external magnetic fields on the spectra of the overlap Dirac operator by calculating the spectral density of the eigenvalues at low and high temperatures. Gaussian random matrix theory (GRMT) predicts fluctuations from short to long ranges of eigenvalues of the Dirac operator universally. Therefore, to investigate the effects of external magnetic fields on the fluctuations of the eigenvalues of the overlap Dirac operator, we analyze the distributions of the nearest-neighbor spacing of the eigenvalues, calculate the spectral rigidity of the eigenvalues, and compare them with the predictions of GRMT~\cite{Dyson2, Guhr2}. In this study, we examine a broader range of the spectral rigidity compared to our previous investigations~\cite{Hasegawa2,Hasegawa4}

Next, we detect the topological charges by analyzing the spectra of the overlap Dirac operator, and use this information to estimate the number of instantons and anti-instantons. The statistics regarding the number of instantons and anti-instantons are insufficient. However, the outcome indicates that their quantity slightly decreases as the intensity of external magnetic fields increases. We then illustrate the effect of external magnetic fields on the distributions of the topological charges. 

In addition, smearing techniques are widely used to generate configurations~\cite{Morningstar_1}, analyze observables, and reduce statistical errors and computational time~\cite{Allton_1,Y_Aoki3,Alexandrou_1,Teper_1}. Calculations of the eigenvalues and eigenvectors of the overlap Dirac operator are time-consuming tasks. To reduce computational time, smearing techniques are used from time to time; however, the smearing effects on observables are unclear. Therefore, we perform smearing to the link variables of the gauge configurations and clarify the smearing effects on monopoles and spectra of the overlap Dirac operator at low and high temperatures. Additionally, at low temperature, we estimate the chiral condensate, which is an order parameter for chiral symmetry breaking, and demonstrate the impact of smearing on chiral symmetry breaking.

Third, we calculate the correlation functions of the connected and disconnected contributions for the pseudoscalar density using eigenvalues and eigenvectors at low temperature. We first investigate the influence of external magnetic field on the Partial Conservation of Axial Currents (PCAC) relation, the decay constant of the pseudoscalar meson, and the chiral condensate.

Next, we extrapolate the decay constants of the pseudoscalar meson and the chiral condensate in the chiral limit. We show that the decay constant in the chiral limit decreases with increasing the magnetic field intensity. The chiral condensate in the chiral limit increases as the magnetic field strength grows. As the intensity of magnetic fields increases, the dimensionless difference between the chiral condensate calculated using configurations applied to external magnetic fields and the chiral condensate calculated using the standard configuration also grows. However, as the chiral condensate is computed as a negative quantity, these outcomes indicate that strengthening external magnetic fields leads to the eventual achievement of zero chiral condensate.

Finally, we determine the square mass of the eta-prime meson using two methods. (1) Using the Witten-Veneziano mass formula and the numerical results. (2) Analyzing the correlation functions of the connected and disconnected contributions for the pseudoscalar density~\cite{DeGrand3,Hashimoto_1}. We extrapolate the numerical results obtained by the second method in the chiral limit and demonstrate that in the chiral limit the square mass of the eta-prime meson becomes lighter by increasing the intensity of external magnetic fields.

This article consists of five sections. Section~\ref{sec:sec2} provides the estimates of the monopole densities, lengths of the monopole loops, and average absolute values of the Polyakov loops.

Section~\ref{sec:sec3} investigates the impacts of external magnetic fields on the spectra of the overlap Dirac operator at low and high temperatures. We compare these spectra with the GRMT predictions and also illustrate how external magnetic fields affect the topological charges along with the number of instantons and anti-instantons. We then demonstrate the smearing effects on the monopoles and spectra of the overlap Dirac operator.

In section~\ref{sec:sec4}, we evaluate the PCAC relation, decay constant, and chiral condensate using the eigenvalues and eigenvectors at low temperature. Finally, we estimate the square masses of the eta-prime meson through two different methods and demonstrate their sensitivity to external magnetic fields.

In section~\ref{sec:sec5}, we provide a summary and conclusion.

Appendix~\ref{sec:sm_finite} contains the results of the analysis of the smearing effects at high temperature.

This article is a contribution to the 3rd international workshop ``Lattice and Functional Techniques for QCD'' held with the 7th international conference ``Models in Quantum Field Theory''. This article presents preliminary outcomes of the initial computations carried out as part of the research cooperative effort. Consequently, we use small lattice volumes ($V = 8^{3}\times16$ and $V = 8^{3}\times4$) of a coarse lattice spacing ($a$ = 0.2457 [fm]). Additionally, the lack of adequate statistics renders our findings inconclusive. However, we found interesting and novel results and report them in this article.

The data tables~\cite{Hasegawa7} supporting this article are accessible on figshare with the digital object identifier: ``\url{https://doi.org/10.6084/m9.figshare.21679526}''. 

%%%%%%%%%%%%%%%%%%%%%%%%%%%%%%%%%%%%%%%%%%%%%%%%%%%%%%%%%%%%%%%%%%%%%%%%%%%
%% SEC 2
%%%%%%%%%%%%%%%%%%%%%%%%%%%%%%%%%%%%%%%%%%%%%%%%%%%%%%%%%%%%%%%%%%%%%%%%%%%

\section{Effects of external magnetic fields on monopoles and Polyakov loop values}\label{sec:sec2}

In this section, we demonstrate the effects of external magnetic fields on the monopoles and Polyakov loop values at low and high temperatures.
 
\subsection{Simulation parameters}\label{sec:1_1}

The Pisa group is conducting studies of how external magnetic fields affect observables at different temperatures~\cite{Bonati3,Bonati6}. In this study, we generate gauge configurations of $N_{f}$ = 2 + 1 flavor quarks with the physical bare quark masses. The standard configurations and the configurations to which external magnetic fields are applied are prepared at low and high temperatures to investigate the impacts of the magnetic fields on observables.

The fermion and gauge actions remain unchanged from their previous study~\cite{Bonati3} and the details of the way of applying external magnetic fields to the configurations are explained in reference~\cite{Massimo1}, which we will briefly outline.

The QCD Lagrangian incorporates external electromagnetic field via the covariant derivative $D_{\mu} = \partial_{\mu} + igA_{\mu}^{a}T^{a}+iq_{f}a_{\mu}$, where $A_{\mu}^{a}$ is the non-Abelian gauge field, $a_{\mu}^{a}$ is the Abelian gauge field, and $q_{f}$ represents the electric charge of the quark flavor $f$. 

In the presence of a magnetic field $B$, the Euclidean partition function can be represented as
\begin{equation}
  \mathcal{Z}(B) = \int\mathcal{D}U\exp{(-S_{\text{YM}})}\prod_{f = u, d, s}\det{(D_{\text{ST}}^{f}[B])}^{\frac{1}{4}}.
\end{equation}
The Symanzik tree-level improved gauge action is given by 
\begin{equation}
  S_{\text{YM}} = -\frac{\beta}{3}\sum_{x}\left(\frac{5}{3}\sum_{\mu > \nu}W_{\mu, \nu}^{1\times1}(x) -\frac{1}{12}\sum_{\mu \neq \nu}W_{\mu, \nu}^{1\times2}(x)  \right).
\end{equation}
The action incorporates the real components of the traces of $1\times1$ loops $W_{\mu, \nu}^{1\times1}(x)$ and $1\times2$ rectangles $W_{\mu, \nu}^{1\times2}(x)$. The definition of the stout-improved staggered Dirac operator is
\begin{equation}
  D_{\text{ST}}^{f}[B] = am_{f}\delta_{x, y} + \sum_{\nu = 1}^{4}\frac{\eta_{\nu}(x)}{2}
  \left(
u_{\nu}^{f}(B, x)U_{\nu}(x)\delta_{x, y - \hat{\nu}} - u_{\nu}^{f*}(B, x - \hat{\nu})U_{\nu}^{\dagger}(x - \hat{\nu})\delta_{x, y + \hat{\nu}}
  \right),
\end{equation}
where $U_{\nu}(x)$ is the two-time stout smeared link in SU(3) which is defined in~\cite{Morningstar_1}, $\eta_{\nu}(x)$ is the staggered phase, and $u_{\nu}^{f}(B, x)$ is the Abelian link variable including external magnetic fields $B$.

The bare quark masses $am_{u,d}$ and $am_{s}$ are tuned to obtain the physical values of the pion mass $m_{\pi} \sim 135$ [MeV] and of the ratio $\frac{m_{s}}{m_{u,d}} = 28.15$~\cite{Borsanyi_1}. In this study, the parameter $\beta$ of the lattice spacing and the bare quark mass $am_{s}$ are determined by interpolating the outcomes in Table 1 in reference~\cite{Borsanyi_1} using cubic splines. The bare quark mass $am_{u,d}$ is obtained through the relation $\frac{m_{s}}{m_{u,d}} = 28.15$~\cite{Borsanyi_1,Y_Aoki1}.
\begin{table}[htbp]
  \begin{center}
    \caption{Simulation parameters of the standard configurations and the configurations to which external magnetic fields are applied. The magnetic field of the standard configuration is mentioned as $Bz$ = 0 and $e|B|$ = 0.00. The systematic errors of the lattice spacing are approximately 2$\%$~\cite{Y_Aoki1}.}\label{tb:lattice}
    \begin{tabular}{|c|c|c|c|c|c|c|c|c|}\hline
      $V$            & $\beta$ & $am_{u,d}$ & $am_{s}$ & $a$ & $T$ & $Bz$ & $e|B|$ &  $N_{\text{conf}}$ \\
      &  & $\times10^{-3}$  & & [fm] & [MeV] &  & [GeV$^{2}$] & \\ \hline
      & & &  &          &           & 0 & 0.00     & 60 \\ \cline{7-9}
      8$^{3}\times$16  & 3.505 & 4.395 & 0.1237 & 0.2457       & 50   & 3            & 0.57 & 60 \\ \cline{7-9}
      & & &  &          &           & 6            & 1.14 & 60 \\ \hline
      & & &  &          &           & 0  & 0.00    & 60 \\ \cline{7-9}
      8$^{3}\times$4  & 3.505 & 4.395 & 0.1237 & 0.2457      & 200     & 3            & 0.57 & 60 \\ \cline{7-9}
      & & &  &           &           & 6            & 1.14 & 60 \\ \hline
    \end{tabular}
  \end{center}
\end{table}

The uniform magnetic fields are applied along the $z$ direction and denoted as $Bz$ in this article. The intensity of these fields is calculated by the following formula:
\begin{equation}
e|B| = \frac{6\pi Bz}{a^{2}L_{x}L_{y}},
\end{equation}
where $L_{x}$ and $L_{y}$ are the lengths of the $x$-direction and $y$-direction of the lattice volume, respectively. We vary the values of $Bz$ from 3 to 6; hence the intensity of external magnetic fields varies from $e|B|$ = 0.57 to 1.14 [GeV$^{2}$]. Hereafter, for convenience, we consider the intensity of the magnetic fields of the normal configuration as zero and indicate it as $Bz = 0$ and $e|B|$ = 0.00.

External magnetic fields do not affect the lattice spacing, as confirmed by simulations~\cite{Bonati6}. In this study, to confirm the interpolated outcome of the lattice spacing, we perform APE~\cite{Ape1} smearing to the spatial components of the non-Abelian link variables and calculate the static potential from the Wilson loop calculations to evaluate the lattice spacing for each magnetic field strength. The numerical results of the lattice spacing $a$ are as follows: $Bz = 0$, $(N, \alpha) = (6, 0.5)$, $a = 0.220(12)$ [fm]. $Bz = 3$, $(N, \alpha) = (6, 0.4)$, $a = 0.220(12)$ [fm]. $Bz = 6$, $(N, \alpha) = (6, 0.5)$, $a = 0.249(13)$ [fm]. $N$ and $\alpha$ indicate the smearing parameters, which represent the number of iteration steps and the weight parameter, respectively.

The calculated results of the lattice spacing indicate consistency. However, the calculated results are approximately 10$\%$ smaller in comparison with the outcome derived from interpolation. We suppose that this discrepancy arises from the finite lattice volume and discretization effects, and we use the lattice spacing value of $a = 0.2457$ [fm] obtained from interpolation in this study.

The configurations are prepared for two distinct temperatures at low temperature $T$ = 50 [MeV] in the confinement phase and at high temperature $T$ = 200 [MeV] in the deconfinement phase. The total number of configurations for each parameter is $N_{\text{conf}}$ = 60. A summary of the simulation parameters can be found in Table~\ref{tb:lattice}.

\subsection{Effects of external magnetic fields on monopoles in color confinement and deconfinement phases}

We transform the SU(3) matrix while preserving the local U(1) $\times$ U(1) invariance under a gauge condition of the maximal Abelian gauge~\cite{Brandstaeter1,Bornyakov4}. This gauge condition maximizes the following function $F[U]$:
\begin{equation}
  F[U] = \frac{1}{12V}\sum_{s, \mu}[|\tilde{U}_{11}(s, \mu)|^{2} + |\tilde{U}_{22}(s, \mu)|^{2} + |\tilde{U}_{33}(s, \mu)|^{2}],
\end{equation}
by performing the local gauge transformation $g$ to the non-Abelian link variable $U(s, \mu)$ as follows:
\begin{equation}
  U(s, \mu) \rightarrow \tilde{U}(s, \mu) = g^{\dagger}(s)U(s, \mu)g(s + \hat{\mu})
\end{equation}
Where $s$ and $\mu$ indicate the site and direction, respectively.

However, if the function $F[U]$ becomes trapped in local maxima rather than true maxima, there may be inaccuracies in the numerical results~\cite{Gribov1}. To solve this problem, we update the link variables of the configurations 20 times for each configuration using the simulated annealing algorithm~\cite{Bali3} and select the configurations with the highest value of $F[U]$. We then calculate the observables using these configurations. The procedure is explained in section ``C. Gribov copies'' in reference~\cite{Bornyakov4}.

The function $F[U]$ preserves the local U(1) $\times$ U(1) invariance. We perform the Abelian projection~\cite{Brandstaeter1} and derive the Abelian component that holds the U(1) $\times$ U(1) symmetry from the non-Abelian link variables. The Abelian component comprises two elements, namely the monopole and photon contributions~\cite{Kronfel1,Polikarpov1}. The monopole current, denoted as $\kappa_{\mu}^{i}$~\cite{DeGrand1}, is defined on the dual sites represented as $^{*}s$ in the following manner:
\begin{equation}
  \kappa_{\mu}^{i} (^{*}s) \equiv - \epsilon_{\mu\nu\rho\sigma}\nabla_{\nu}\theta_{\rho\sigma}^{i}(s + \hat{\mu})\label{eq:mon_curr},
\end{equation}
where the superscript $i$ indicates the color. The forward derivative defined on the lattice is indicated by the derivative $\nabla_{\nu}$. The right-hand side of~(\ref{eq:mon_curr}) represents the number of Dirac strings that enter and leave the regular hexahedrons in three dimensions, which are composed of $1\times 1$ plaquettes. The monopole density $\rho_{m}$ is calculated according to the method described in~\cite{Bornyakov4} as follows:
\begin{equation}
  a^{3}\rho_{m} = \frac{1}{12V}\sum_{i, \mu}\sum_{^{*}s}|\kappa_{\mu}^{i}(^{*}s)|
\end{equation}
The numerical results of the monopole density are listed in Table~\ref{tb:Ploops_vs_mloops}.
 \begin{table}[htbp]
    \begin{center}
        \caption{The computed outcomes include the monopole density $\rho_{m}$, the lengths of the longest and rest monopole loops L$^{\text{long}}$ and L$^{\text{rest}}$, and the mean absolute value of the Polyakov loops $\langle |P| \rangle$.}\label{tb:Ploops_vs_mloops}
 \begin{tabular}{|c|c|c|c|c|c|}\hline
   $V$ & $Bz$ & $\rho_{m}$ [GeV$^{3}$] & L$^{\text{long}}$ [fm] & L$^{\text{rest}}$ [fm] & $\langle |P| \rangle$ \\ \hline
           &  0 & 5.24(3)$\times10^{-2}$ & 185.4(1.2) & 18.2(3) & 1.22(9)$\times10^{-2}$  \\ \cline{2-6}
   8$^{3}\times$16 & 3 & 5.48(2)$\times10^{-2}$ & 196.3(1.1) & 16.7(2) & 1.20(9)$\times10^{-2}$  \\ \cline{2-6}
                   &   6 & 5.70(3)$\times10^{-2}$ & 205.9(1.1) & 15.7(2) & 1.27(9)$\times10^{-2}$  \\ \hline
          & 0 & 4.16(5)$\times10^{-2}$ & 28.1(9) & 12.3(5) & 0.116(2) \\ \cline{2-6}
   8$^{3}\times$4 & 3 & 4.02(9)$\times10^{-2}$ & 25.0(1.1) & 14.0(6) & 0.120(2) \\ \cline{2-6}
                   & 6 &  3.95(10)$\times10^{-2}$ &23.6(9) & 14.7(6) & 0.134(2) \\ \hline   
 \end{tabular}
  \end{center}
 \end{table}

The monopole current preserves the following current conservation law: $\nabla_{\mu}^{*}k_{\mu}^{i} (^{*}s) = 0$, where the derivative $\nabla_{\mu}^{*}$ indicates the backward derivative. Therefore, the monopole currents form closed loops. We count the number of monopole currents composing the closed loops using a calculation method~\cite{Bode1} and measure the length of the monopole loops $L$, which is defined as follows:
 \begin{equation}
  L/a \equiv \frac{1}{12}\sum_{i, \mu}\sum_{^{*}s \in C}|\kappa_{\mu}^{i}(^{*}s)|
\end{equation}
\begin{figure*}[htbp]
  \begin{center}
    \includegraphics[width=155mm]{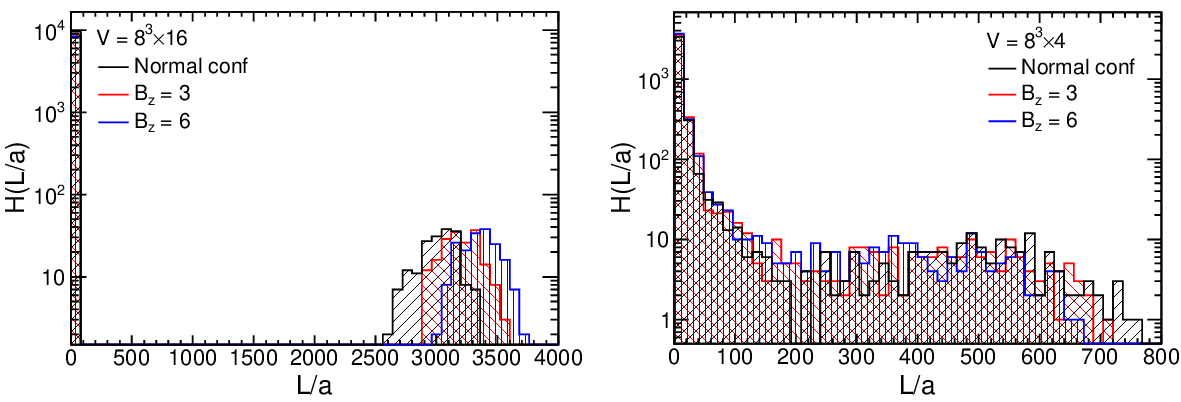}
  \end{center}
  \setlength\abovecaptionskip{-1pt}
  \caption{Histograms H($L/a$) displaying the length of the monopole loops $L/a$ at low (left) and high (right) temperatures.}\label{fig:hist_mloops}
\end{figure*}

First, we create histograms H($L/a$) of the monopole loops $L/a$ at low and high temperatures. The left panel of Fig.~\ref{fig:hist_mloops} indicates that the monopole loops at low temperature are divided into two clusters - one consisting of short loops and the other consisting of long loops. At low temperatures, there is one long monopole loop for each color component. The longest monopole loops are closely related to the color confinement mechanism~\cite{Kitahara1,Kronfel2}, and the length correlates with the transition temperature from the confinement phase to the deconfinement phase~\cite{Hasegawa4} and the number of instantons and anti-instantons~\cite{DiGH4}.

However, the right panel of Fig.~\ref{fig:hist_mloops} demonstrates that, at high temperature, the lengths of the monopole loops cannot be categorized and that the longest monopole loops become shorter compared to those at low temperature.

To demonstrate the effect of external magnetic fields on the monopoles, we measure the length of the longest monopole loops and classify them into two groups: (i) longest monopole loops denoted as $L_{m}^{\text{long}}$, and (ii) the remaining loops denoted as $L_{m}^{\text{rest}}$. The rest loops are calculated by subtracting the longest loops from the total number of loops. The physical lengths of the monopole loops $L_{m}^{\text{long}}$ and $L_{m}^{\text{rest}}$ are represented in Table~\ref{tb:Ploops_vs_mloops}.

Figure~\ref{fig:hist_mloops} and the outcomes in Table~\ref{tb:Ploops_vs_mloops} demonstrate the following effects of external magnetic fields on the monopoles at low and high temperatures.\\

\noindent At low temperature:
\begin{itemize}
\item The monopole density increases with the intensity of external magnetic fields.
\item The length of the longest monopole loops increases, whereas the length of the remaining loops decreases as the intensity of external magnetic fields increases.
\end{itemize}
At high temperature:
\begin{itemize}
\item The monopole density decreases with increasing the intensity of external magnetic fields.
\item The length of the longest monopole loops decreases, whereas the length of the remaining loops increases as the intensity of external magnetic fields increases.
\end{itemize}

This study utilizes the configurations with the dynamical fermions. As such, the average absolute value of the Polyakov loops $\langle |P| \rangle$ cannot serve as an order parameter for the phase transition of color deconfinement. Nonetheless, we calculate the average absolute values of the Polyakov loops and compare them with the lengths of the longest monopole loops to demonstrate the effects of external magnetic fields at low and high temperatures.
\begin{figure*}[htbp]
  \begin{center}
    \includegraphics[width=150mm]{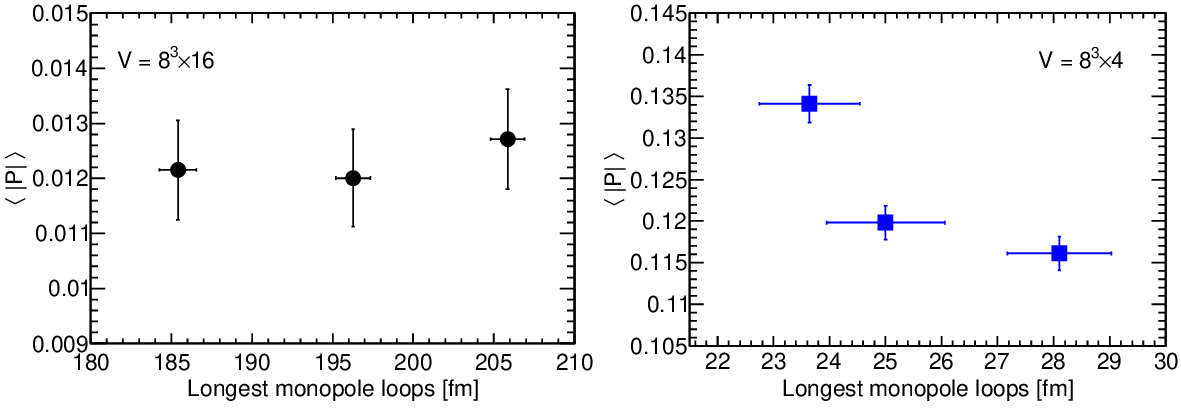}
  \end{center}
  \setlength\abovecaptionskip{-1pt}
  \caption{Comparisons of the length of the longest monopole loops $L_{m}^{\text{long}}$ and the average absolute value $\langle |P| \rangle$ of the Polyakov loops. The left and right panels show the outcomes at low ($V = 8^{3}\times16$) and high temperatures ($V = 8^{3}\times4$), respectively.}\label{fig:ploops_vs_mloops}
\end{figure*}

The Polyakov loop is defined as follows:
\begin{equation}
  P(\vec{s}) = \text{Tr}\prod_{s_{0} = 0}^{N_{t}-1}U(s, \hat{0}),
\end{equation}
where $U(s, \hat{\mu})$ denotes the SU(3) non-Abelian link variable. The average absolute values $\langle |P| \rangle$ of the Polyakov loops $P$ are represented in Table~\ref{tb:Ploops_vs_mloops}.

Figure~\ref{fig:ploops_vs_mloops} shows that at low temperature the longest monopole loops become longer when the intensity of external magnetic fields becomes strong. However, external magnetic fields do not affect the average absolute value of the Polyakov loops. At high temperature, the longest monopole loops become shorter, and the average absolute values of the Polyakov loops increase when the intensity of external magnetic fields becomes strong.

After section~\ref{sec:sec2}, we do not transform the SU(3) matrices under a certain gauge condition nor perform the Abelian projection to the SU(3) matrices.

%%%%%%%%%%%%%%%%%%%%%%%%%%%%%%%%%%%%%%%%%%%%%%%%%%%%%%%%%%%%%%%%%%%%%%%%%%%
%% SEC 3
%%%%%%%%%%%%%%%%%%%%%%%%%%%%%%%%%%%%%%%%%%%%%%%%%%%%%%%%%%%%%%%%%%%%%%%%%%%

\section{Spectra of the overlap Dirac operator, topological charges, and instantons in external magnetic fields}\label{sec:sec3}

In this section, we calculate the eigenvalues and eigenvectors of the overlap Dirac operator and analyze the influence of external magnetic fields on the spectra of the Dirac operator.

\subsection{Spectra of the overlap Dirac operator}\label{sec:spect_ov}

The Dirac operator $\mathcal{D}$ that preserves exact chiral symmetry in the continuum limit in lattice gauge theory satisfies the following Ginsparg-Wilson relation~\cite{Ginsparg1}:
\begin{equation}
\gamma_{5}\mathcal{D} + \mathcal{D} \gamma_{5} = a\mathcal{DR}\gamma_{5}\mathcal{D},
\end{equation}
where $\gamma_{5}$ is the fifth component of the Dirac gamma matrices in the Dirac basis and $\mathcal{R}$ is a local operator.

The massless overlap Dirac operator $\mathcal{D}$ that satisfies the Ginsparg-Wilson relation is derived as follows~\cite{Neuberger1,Neuberger2,Lusher1}\footnote{We use the same notation as in reference~\cite{Galletly_1}.}:
\begin{align}
  \mathcal{D}(\rho) = \frac{\rho}{a} \left[ 1 + \frac{\mathcal{D}_{W}(\rho)}{ \sqrt{\mathcal{D}_{W}^{\dagger}(\rho)\mathcal{D}_{W}(\rho)}} \right], \ \ \mathcal{D}_{W}(\rho) = \mathcal{D}_{W} - \frac{\rho}{a}\label{eq:overlap_d},
\end{align}
where $\mathcal{D}_{W}$ is the standard Wilson-Dirac operator. We set the negative mass parameter $\rho$ to $\rho = 1.4$. This massless overlap Dirac operator is approximated using the sign function $\text{sign}\left\{\mathcal{H}_{W}(\rho)\right\}$ of the Hermitian Wilson-Dirac operator $\mathcal{H}_{W}(\rho)$ as follows:
\begin{align}
  \mathcal{D}(\rho) = \frac{\rho}{a}\left[1 + \gamma_{5}\text{sign}\left\{\mathcal{H}_{W}(\rho)\right\}\right], \ \ \text{sign}\left\{\mathcal{H}_{W}(\rho)\right\} = \frac{\mathcal{H}_{W}(\rho)}{\sqrt{\mathcal{H}_{W}^{\dagger}(\rho)\mathcal{H}_{W}(\rho)}}\label{eq:overlap_d_sign}
\end{align}
The Hermitian Wilson-Dirac operator is defined as $\mathcal{H}_{W}(\rho) = \gamma_{5}\mathcal{D}_{W}(\rho)$.

The sign function $\text{sign}\left\{\mathcal{H}_{W}(\rho)\right\}$ is calculated using the (minmax) polynomial approximation $P_{n, \epsilon}$ of the degree $n$ of the interval [$\epsilon$, 1] as follows: $\text{sign}\left\{\mathcal{H}_{W}(\rho)\right\} = HP_{n, \epsilon}(H^{2}), \ \ H = \frac{\mathcal{H}_{W}(\rho)}{||\mathcal{H}_{W}(\rho)||}$. The parameter $\epsilon$ is determined from the following formula: $\epsilon = \left(\frac{\lambda_{\text{min}}}{\lambda_{\text{max}}}\right)^{2}$, where $\lambda_{\text{min}}$ and $\lambda_{\text{max}}$ are the minimum and maximum eigenvalues of the Hermitian Wilson-Dirac operator, respectively. The degree of the polynomial function $n$ is determined for each configuration. We use the low-mode projector, which is a numerical technique to calculate the sign function using the eigenvalue decompositions. Reference~\cite{Giusti6} explains the details of the numerical calculations.
\begin{figure*}[htbp]
  \begin{center}
    \includegraphics[width=75mm]{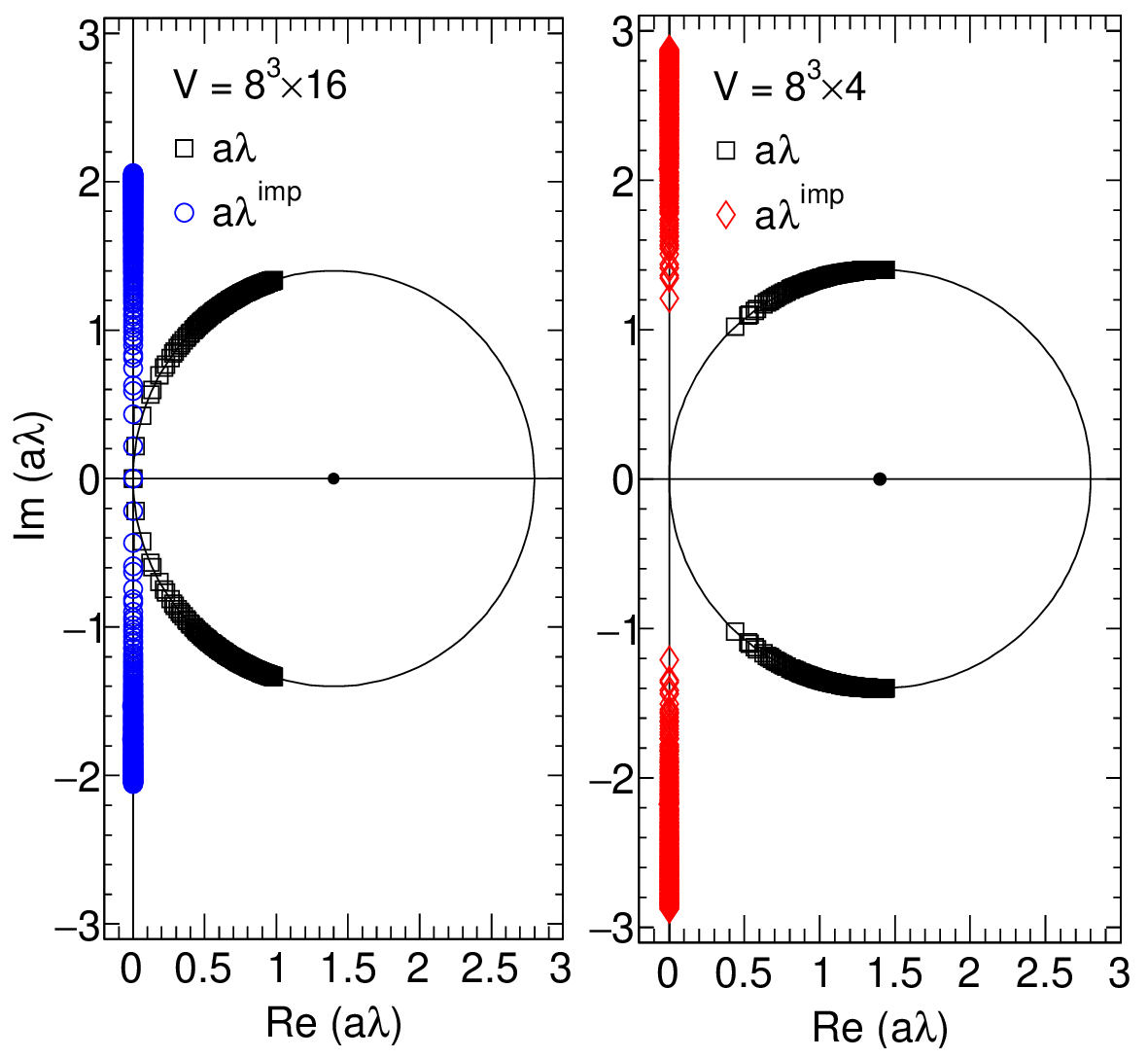}
  \end{center}
  \setlength\abovecaptionskip{-1pt}
  \caption{The eigenvalues $a\lambda$ and improved eigenvalues $a\lambda^{\text{imp}}$ at low (left) and high (right) temperatures.}\label{fig:Spec_8xx3x16_8xx3x4}
\end{figure*}

We solve the following eigenvalue problem using the ARPACK subroutine: $\mathcal{D}| \psi_{i} \rangle = \lambda_{i} | \psi_{i} \rangle$, where $\lambda_{i}$ and $\psi_{i}$ are the eigenvalues and eigenvectors of the massless overlap Dirac operator~(\ref{eq:overlap_d}), respectively. The subscript $i$ indicates the mode number. We calculate pairs of eigenvalues and eigenvectors from the lowest energy level to approximately 400 for each configuration and save them to the storage elements.

Fermion zero modes appear in the spectra of the overlap Dirac operator. The eigenvalues of the overlap Dirac operator are located on a circle in the complex plane, and their complex conjugates always exist. The radius of this circle in this study is 1.4, based on the mass parameter of $\rho$ = 1.4, with the circle centered at (Re, Im) = (1.4, 0).

We compute the improved eigenvalues $a\lambda^{\text{imp}}$ of the following improved Dirac operator $\mathcal{D}^{imp}$~\cite{Capitani}:
\begin{equation}
\mathcal{D}^{imp} = \left( 1 - \frac{a}{2\rho} \mathcal{D} \right)^{-1} \mathcal{D}\label{imp_op1}
\end{equation}
to obtain eigenvalues near the continuum limit. The improved eigenvalues $a\lambda^{\text{imp}}$ are projected onto the imaginary axis, resulting in pure imaginary numbers. We present Fig.~\ref{fig:Spec_8xx3x16_8xx3x4}, illustrating examples of the spectra $a\lambda$ and $a\lambda^{\text{imp}}$ at low and high temperatures.

The Banks-Casher relation suggests that the breaking of chiral symmetry occurs spontaneously due to a nonzero density of eigenvalues near the energy level of zero~\cite{Banks1}. Thus, to illustrate the effect of external magnetic fields on chiral symmetry breaking, our initial step involves computing the spectral density which is defined as follows:
\begin{equation}
    \rho (\lambda) = \frac{1}{V}\left\langle \sum_{\lambda}\delta(\lambda - \bar{\lambda})\right\rangle\label{eq:spec_1}
  \end{equation}
\begin{figure*}[htbp]
  \begin{center}
    \includegraphics[width=160mm]{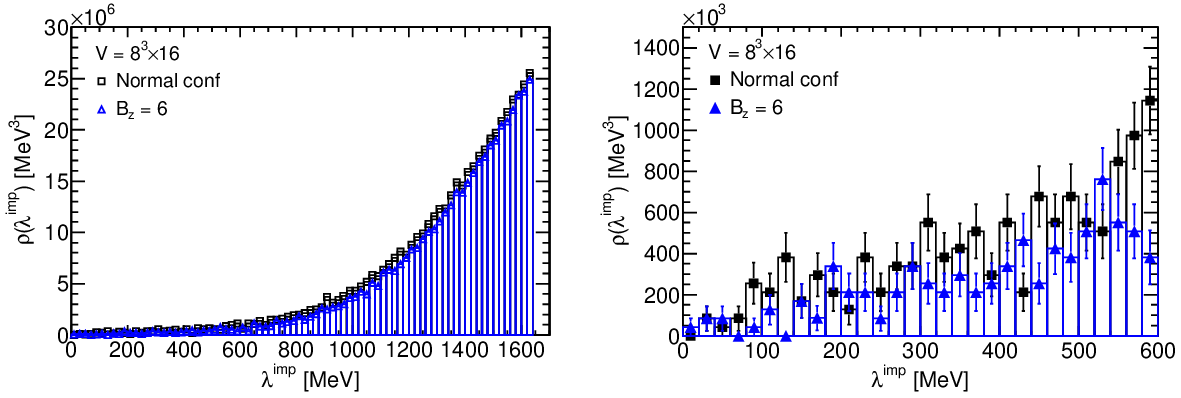}
  \end{center}
  \setlength\abovecaptionskip{-1pt}
  \caption{The spectral density $\rho(\lambda^{\text{imp}})$ of the improved eigenvalues $\lambda^{\text{imp}}$ at low temperature ($V = 8^{3}\times16$). The right panel shows the spectral density in the range from 0 to 600 [MeV].}\label{fig:Spec_dens_8xx3x16}
\end{figure*}

Here, for clarity, we express the improved eigenvalues of the pure imaginary numbers in the lattice unit as $\pm ia\lambda_{k}^{\text{imp}}$. In this study, we select the improved eigenvalues of the positive values excluding the zero eigenvalues as $a\lambda_{k}^{\text{imp}}$ and arrange the eigenvalues in ascending order. The subscript $k$ denotes the new mode number of positive and nonzero eigenvalues, ranging from $k$ = 1 to 190. We count the number of improved eigenvalues at intervals of 20 [MeV] and calculate the spectral density (\ref{eq:spec_1}). Figure \ref{fig:Spec_dens_8xx3x16} illustrates that the spectral density at low temperature decreases slightly as the intensity of external magnetic fields increases.

We have calculated numerically the overlap Dirac operator at high temperatures in quenched QCD. We set the temperature to approximately the same value as the full QCD case with $\frac{T}{T_{c}} = 1.32$. The lattice is $V$ = $26^{3}\times6$ of $\beta$ = 6.059 ($a$ = 8.435$\times10^{-2}$ [fm]), the physical lattice volume of this study ($V = 8^{3}\times4$) is approximately 1.4 times larger than the physical volume of quenched QCD.
\begin{figure*}[htbp]
  \begin{center}
    \includegraphics[width=160mm]{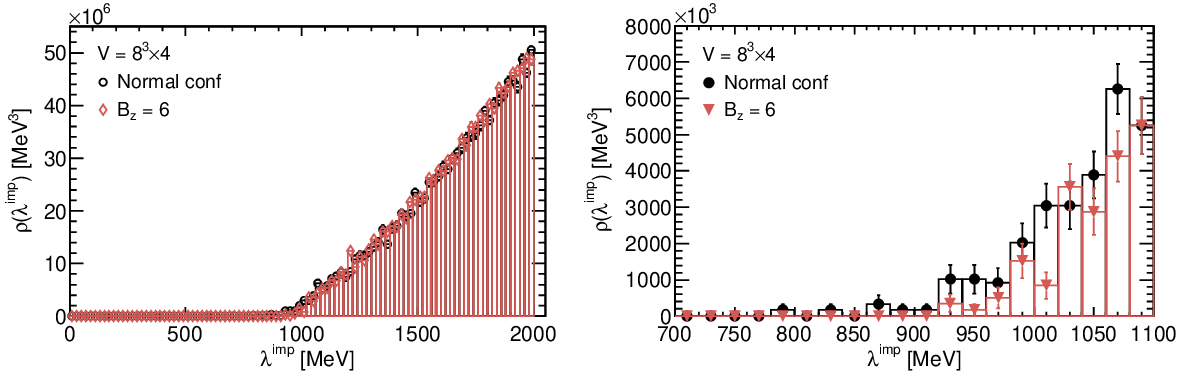}
  \end{center}
  \setlength\abovecaptionskip{-1pt}
  \caption{The spectral density $\rho(\lambda^{\text{imp}})$ of the improved eigenvalues $\lambda^{\text{imp}}$ at high temperature ($V = 8^{3}\times4$). The right panel shows the spectral density in the range from 700 to 1100 [MeV].}\label{fig:Spec_dens_8xx3x4}
\end{figure*}
\begin{figure*}[htbp]
  \begin{center}
    \includegraphics[width=160mm]{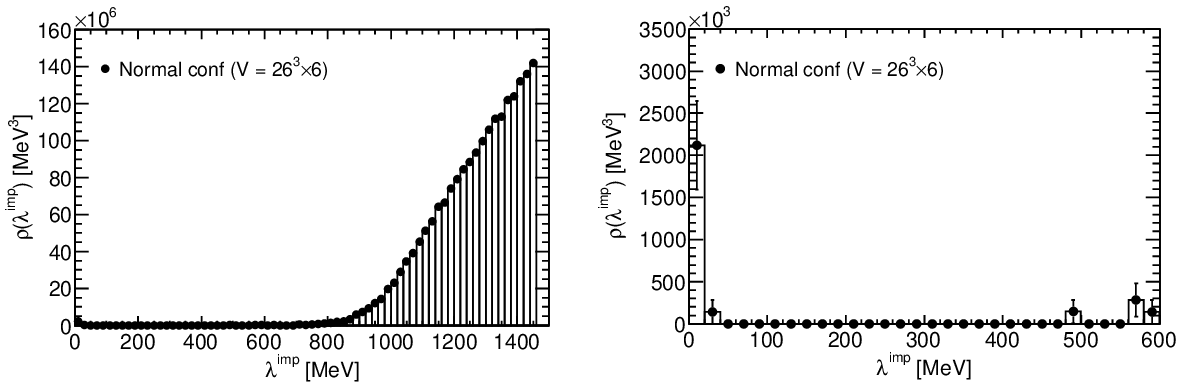}
  \end{center}
  \setlength\abovecaptionskip{-1pt}
  \caption{The spectral density $\rho(\lambda^{\text{imp}})$ of quenched QCD at high temperature ($\frac{T}{T_{c}}$ = 1.32, $V = 26^{3}\times6$) is presented using the improved eigenvalues $\lambda^{\text{imp}}$. The right panel displays the spectral density ranging from 0 to 600 [MeV].}\label{fig:Spec_dens_26xx3x6}
\end{figure*}

Figure~\ref{fig:Spec_dens_8xx3x4} shows that the low modes near zero energy level, which are observed in the studies of quenched QCD as depicted in Fig.~\ref{fig:Spec_dens_26xx3x6}, are not present in studies of full QCD. Moreover, Figure~\ref{fig:Spec_dens_8xx3x4} shows that the spectral density decreases at high temperature only near the 900 [MeV] gap, and increasing the intensity of external magnetic fields has no effect on the higher modes.

\subsection{Comparisons of the spectra with GRMT}
  
Fluctuations of the eigenvalues of the Dirac operator are universally predicted by Gaussian Random Matrix Theory (GRMT)~\cite{Dyson2,Guhr2}. There are three canonical ensembles that result from symmetries: the Gaussian orthogonal ensemble, the Gaussian unitary ensemble, and the Gaussian symplectic ensemble. We compare the spectra of the overlap Dirac operator with the predictions of GRMT to investigate the effects of external magnetic fields.

To compare the spectra, we first compute the unfolded eigenvalues using the methods described in ``2. Configuration unfolding'' in reference~\cite{Guhr1}. We plot the improved eigenvalues $a\lambda_{k}^{\text{imp}}$ obtained in subsection~\ref{sec:spect_ov} on the x-axis and the mode number $k$ on the y-axis. Subsequently, we fit the following polynomial function of the fourth degree to the data and obtain the fitting parameters $p_{d}^{n}$, ($d = 0-4$):
\begin{equation}
N_{pol}^{n}(\lambda_{n}) = p_{0}^{n}\lambda_{n}^{0} + p_{1}^{n}\lambda_{n}^{1} + p_{2}^{n}\lambda_{n}^{2} + p_{3}^{n}\lambda_{n}^{3} + p_{4}^{n}\lambda_{n}^{4}.
\end{equation}
The superscript $n$ is the configuration number. We perform this procedure for the numerical results of each configuration. We then substitute the improved eigenvalues of each configuration into the following formula and obtain the unfolded eigenvalues $\xi_{k}^{n}$:
\begin{equation}
  \xi_{k}^{n} = N_{pol}^{n}(a\lambda_{k}^{n}).
\end{equation}
\begin{figure*}[htbp]
  \begin{center}
    \includegraphics[width=155mm]{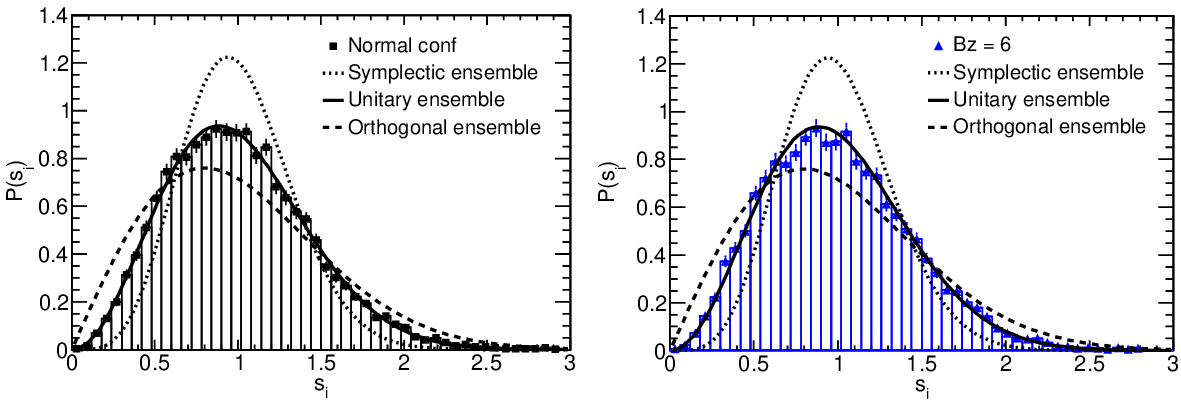}
  \end{center}
  \setlength\abovecaptionskip{-1pt}
  \caption{The distributions P($s_{i}$) of the nearest-neighbor spacing $s_{i}$ at low temperature ($V = 8^{3}\times16$). The distributions of the normal configurations (left) and the magnetic field $B_{z}$ = 6 (right).}\label{fig:near}
\end{figure*}
\begin{figure*}[htbp]
  \begin{center}
    \includegraphics[width=155mm]{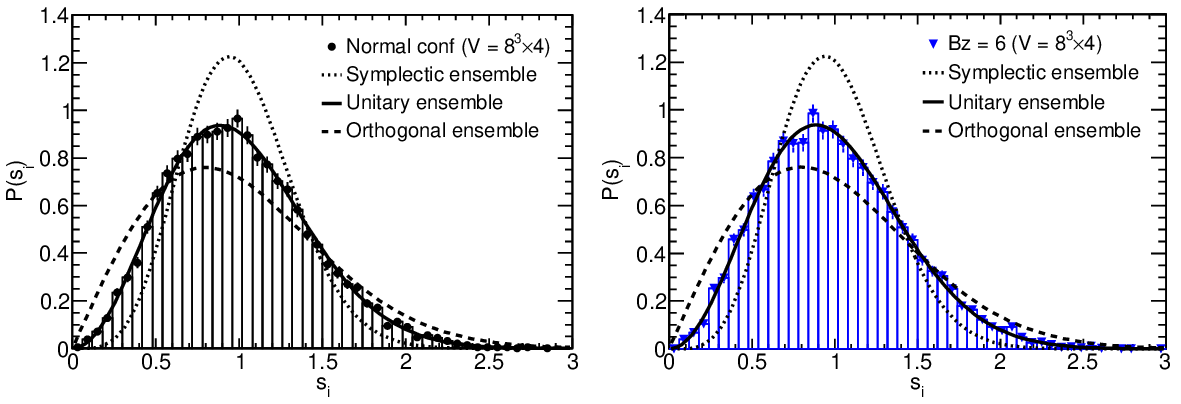}
  \end{center}
  \setlength\abovecaptionskip{-1pt}
  \caption{The distributions P($s_{i}$) of the nearest-neighbor spacing $s_{i}$ at high temperature ($V = 8^{3}\times4$). The distributions of the normal configurations (left) and the magnetic field $B_{z}$ = 6 (right).}\label{fig:near_8xx3x4}
\end{figure*}

To investigate the effect of external magnetic fields on short-range fluctuations, we initially compute the nearest-neighbor spacing using the unfolded eigenvalues denoted as $s_{k}^{n}$. This quantity is defined as follows:  
\begin{equation}
  s_{k}^{n} = \xi_{k+1}^{n} - \xi_{k}^{n}\label{eq:near_si}.
\end{equation}
Subsequently, we make the normalized histogram P($s_{i}$) of the nearest-neighbor spacing $s_{i}$. GRMT predicts the distributions of the nearest-neighbor spacing for each ensemble~\cite{Guhr2}; therefore, we compare the distributions with our outcomes. We find that the distributions of the nearest-neighbor spacing match the predicted distribution of the Gaussian unitary ensemble at low and high temperatures, and the distribution of the nearest-neighbor spacing remains unaffected by external magnetic fields, as demonstrated in Figs.~\ref{fig:near} and~\ref{fig:near_8xx3x4}.

Next, we aim to analyze the impact of external magnetic fields on the fluctuations of the long-range interval $L$. The spectral rigidity is defined as the least-squares approximation obtained by fitting a linear function to the staircase function in the interval $L$, such that the mean square deviations are minimized~\cite{Dyson2}.
\begin{figure*}[htbp]
  \begin{center}
    \includegraphics[width=155mm]{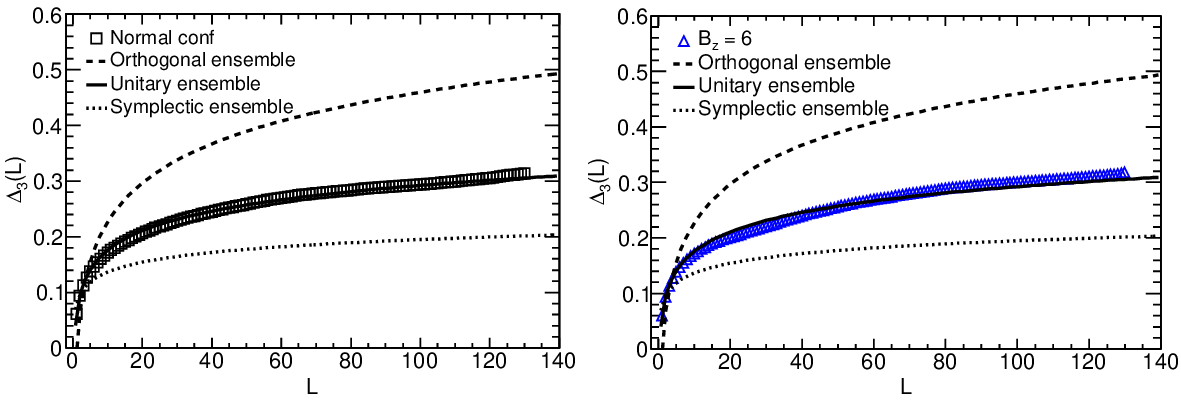}
  \end{center}
  \setlength\abovecaptionskip{-1pt}
  \caption{The comparisons of the spectral rigidity $\Delta_{3}(L)$ at low temperature ($V = 8^{3}\times16$) with the predicted functions of GRMT. The calculated results of the normal configuration (left) and the magnetic field $B_{z}$ = 6 (right).}\label{fig:delta3}
\end{figure*}
\begin{figure*}[htbp]
  \begin{center}
    \includegraphics[width=155mm]{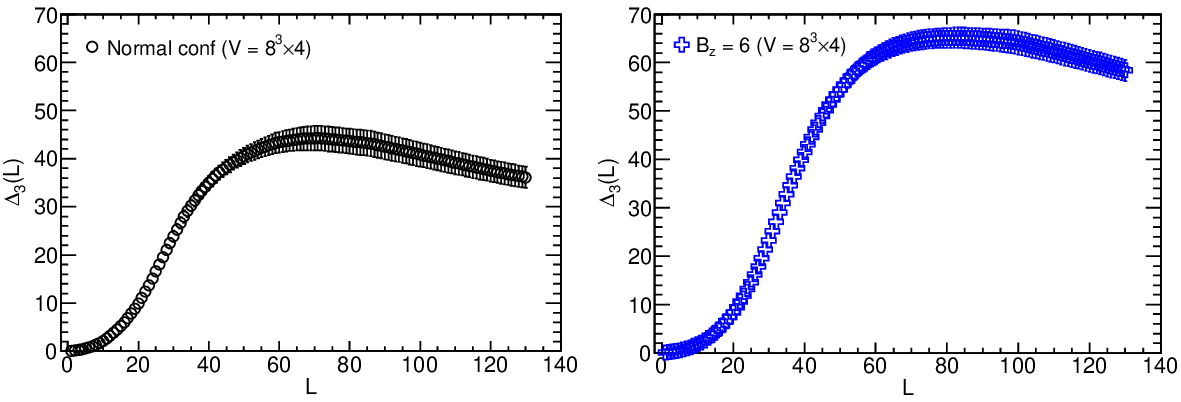}
  \end{center}
  \setlength\abovecaptionskip{-1pt}
  \caption{The calculated results of the spectral rigidity $\Delta_{3}(L)$ at high temperature ($V = 8^{3}\times4$). The calculated results of the normal configuration (left) and the magnetic field $B_{z}$ = 6 (right).}\label{fig:delta3_8xx3x4}
\end{figure*}

We measure the spectral rigidity $\Delta_{3}(L)$ of the specific interval [$\alpha$, $\alpha + L$] on the unfolded scale, with $\alpha$ being the starting point of the measurement, using the analytical formula (I-39) provided in~\cite{Bohigas2_2}. Our first step is to shift the unfolded eigenvalues,
\begin{equation}
  \tilde{\xi}_{k}^{n} = \xi_{k}^{n} - \left(\alpha - \frac{L}{2} \right)
\end{equation}
and then substitute them into the following formula to calculate the spectral rigidity:
\begin{align}
  & \overline{\langle\Delta_{3} (L) \rangle}  = \frac{j^{2}}{16} - \frac{1}{L^{2}}\left(\sum_{k = 1}^{j}\tilde{\xi}_{k}\right)^{2}\nonumber \\
  & \hspace{18mm} + \frac{3j}{2L^{2}}\left(\sum_{k = 1}^{j}\tilde{\xi}_{k}^{2}\right)  - \frac{3}{L^{4}}\left(\sum_{k = 1}^{j}\tilde{\xi}_{k}^{2}\right)^{2}  +\frac{1}{L}\left[\sum_{k = 1}^{j}(j - 2k + 1)\tilde{\xi}_{k}\right]\label{eq:Delta3_rmt}.
\end{align}
The configuration number $n$ is eliminated in the formula. The maximum interval on the unfolded scale is 130. We vary the starting point $\alpha$ on the unfolded scale from 25 to 60, calculate the spectral average, and then evaluate the ensemble average using the Jackknife method.

Figure~\ref{fig:delta3} demonstrates that, at low temperature, the numerical findings conform to the predicted function of the Gaussian unitary ensemble, and that fluctuations of the long-range interval remain unaffected by external magnetic fields. However, Figure~\ref{fig:delta3_8xx3x4} shows that at high temperature the numerical results do not align with the predicted functions of GRMT and that external magnetic fields impact the fluctuations of the long-range interval.

\subsection{Topological charges and instantons}~\label{sec:top_ins}

The number of zero modes of the positive chirality is $n_{+}$ and the number of zero modes of the negative chirality is $n_{-}$. In this study, we find eigenvalues of which the absolute values are less than $1.0\times10^{-7}$ and define them as zero modes. We determine their chirality by multiplying $\gamma_{5}$ by their eigenvectors.

Suppose the Atiyah-Singer index theorem is satisfied~\cite{Atiyah1,Atiyah2}. Let the number of instantons of the positive charge be $n_{+}$ and the number of ant-instantons of the negative charge be $n_{-}$. However, we have never simultaneously observed zero modes of positive and negative chiralities from the same configuration. The zero modes that we detected are topological charges $Q$, which are defined as follows: 
\begin{equation}
Q = n_{+} - n_{-}.
\end{equation}
We have demonstrated that the number of instantons and anti-instantons $N_{I}$ can be approximated from the mean squares of the topological charge~\cite{DiGH3} as follows:
\begin{equation}
  N_{I} = \langle Q^{2} \rangle
\end{equation}
According to this formula, the density of instantons and anti-instantons is in conformity with the topological susceptibility as follows:
\begin{equation}
  \frac{N_{I}}{V} = \frac{\langle Q^{2} \rangle}{V}\label{eq:top_sus}
\end{equation}
Table~\ref{tb:top_ins} displays the numerical results for the absolute value of the topological charges $\langle|Q|\rangle$, the number of instantons and anti-instantons $N_{I}$, the number density of instantons and anti-instantons $\frac{N_{I}}{V}$, and the topological susceptibility $\left(\frac{\langle Q^{2}\rangle}{V}\right)^{\frac{1}{4}}$. Here $\langle \cdots \rangle$ denotes mean values computed from the number of configurations.
 \begin{table}[htbp]
    \begin{center}
        \caption{The calculation results of low temperature include the absolute value of the topological charges denoted as $\langle|Q|\rangle$, the number of instantons and anti-instantons referred to as $N_{I}$, the number density of instantons and anti-instantons represented by $\frac{N_{I}}{V}$, and the topological susceptibility, which is expressed as $\left(\frac{\langle Q^{2}\rangle}{V}\right)^{\frac{1}{4}}$.}\label{tb:top_ins}
        \begin{tabular}{|c|c|c|c|c|c|}\hline
   $V$ & $Bz$ & $\langle|Q|\rangle$ & $N_{I}$ & $\frac{N_{I}}{V}$ [GeV$^{4}$] & $\left(\frac{\langle Q^{2}\rangle}{V}\right)^{\frac{1}{4}}$ [MeV] \\ \hline
           &  0 & 1.20(10) & 2.0(3) & 1.02(14)$\times10^{-4}$ & 1.00(4)$\times10^{2}$ \\ \cline{2-6}
   8$^{3}\times$16 & 3 & 1.07(12) & 2.0(4) & 1.00(18)$\times10^{-4}$ & 1.00(5)$\times10^{2}$ \\ \cline{2-6}
                   &   6 &  0.95(12) & 1.7(3) & 8.7(15)$\times10^{-5}$ & 97(4)  \\ \hline
 \end{tabular}
  \end{center}
 \end{table}
 
The numerical results demonstrate the chiral magnetic effects as seen in the decrease in absolute values of the topological charges $\langle|Q|\rangle$ and the reduction in the number of instantons and anti-instantons $N_{I}$ with increasing the intensity of external magnetic fields. However, the number of instantons and anti-instantons is estimated from the square of the topological charges. As a result, statistical errors are still significant to observe a clear decrease.

\begin{figure*}[htbp]
  \begin{center}
    \includegraphics[width=150mm]{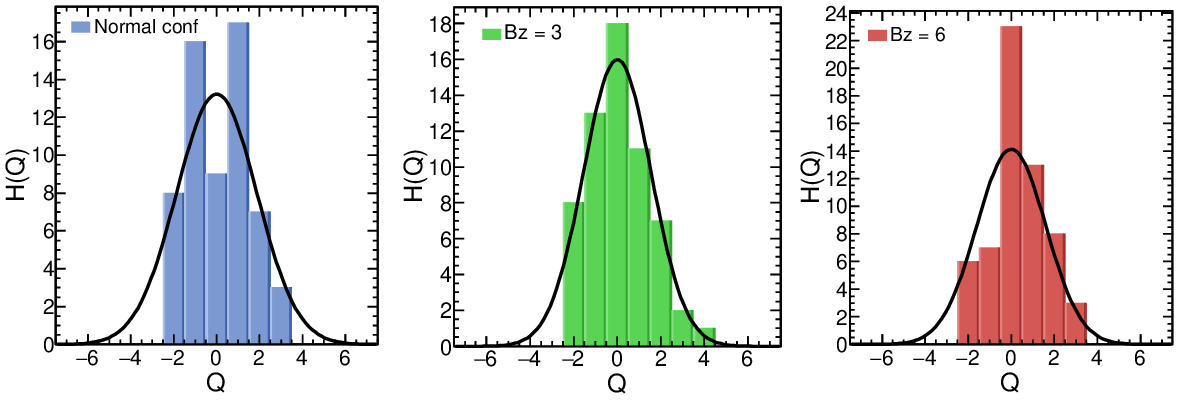}
  \end{center}
  \setlength\abovecaptionskip{-1pt}
  \caption{Histograms H($Q$) of the topological charges $Q$ at low temperature ($V = 8^{3}\times16$). The fitting results by the Gaussian function~(\ref{eq:gauss_1}) are depicted by black curves.}\label{fig:hist_Q}
\end{figure*}

Additionally, Figure~\ref{fig:hist_Q} illustrates the histograms H($Q$) of the topological charges $Q$ computed from the configurations at low temperature. We apply the following Gaussian function, which resulted from the examinations of quenched QCD by other research groups~\cite{Giusti4,DeDebbio1}, to numerical findings:
\begin{equation} 
   P(Q) = \frac{\mathrm{e}^{-\frac{Q^{2}}{2\langle Q^{2}\rangle}}}{\sqrt{2\pi\langle Q^{2}\rangle}}\left[ 1 + \mathcal{O}(V^{-1})\right]\label{eq:gauss_1}.
\end{equation}
\begin{table}[htbp]
  \begin{center}
    \caption{The fitting results of the histograms H($Q$) of the topological charges $Q$ by the Gaussian function~(\ref{eq:gauss_1}).}\label{tb:fit_top}
    \begin{tabular}{|c|c|c|c|c|}\hline
      $V$ & $Bz$ & $\langle Q^{2}\rangle$ & $O(V^{-1})$ & $\chi^{2}/\text{dof}$ \\ \hline
      &  0 & 3.5(1.1) & 0.03(0.16) & 5/4  \\ \cline{2-5}
      8$^{3}\times$16 & 3 & 2.4(0.6) & 0.04(0.14) & 1/5  \\ \cline{2-5}
      &   6 & 2.6(1.0) & -0.05(0.14) & 7/4  \\ \hline
    \end{tabular}
  \end{center}
\end{table}

The fitting results of the correction term $O(V^{-1})$ are zero, and the values of $\chi^{2}/\text{dof}$ are small enough as shown in Table~\ref{tb:fit_top}. Therefore, the fitting function properly approximates the distributions of the topological charges. The fitting results $\langle Q^{2} \rangle$ indicate the number of instantons and anti-instantons. The errors of the fitting results are more than 25$\%$; however, the results indicate that the number of instantons and anti-instantons slightly decreases with increasing the intensity of external magnetic fields.

At high temperature, we observe one topological charge $|Q|$ = 1 in 60 configurations of each type, and the topological charges of the remaining configurations are $|Q|$ = 0. The numerical findings of quenched QCD for the lattice $V$ = $26^{3}\times6$ of $\beta$ = 6.059 indicate that $\langle|Q|\rangle$ = 0.60(5), $N_{I}$ = 0.87(12), $\frac{N_{I}}{V}$ = 2.5(3)$\times10^{-4}$ [GeV$^{4}$], $\frac{\langle Q^{2} \rangle}{V}$ = $(125 \pm 4)^{4}$ [MeV$^{4}$], and $N_{\text{conf}}$ = 201. These results suggest that the configurations with $N_{f}$ = 2 + 1 quark flavors effectively suppress the creation of the topological charges or zero modes. 

From the numerical results in Table~\ref{tb:top_ins}, the topological susceptibility~(\ref{eq:top_sus}) using the normal configurations at low temperature is $\frac{\langle Q^{2} \rangle}{V}$ = $(100 \pm 4)^{4}$ [MeV$^{4}$]. The result of another study~\cite{Bonati5} that utilizes the same fermion and gauge actions as this study is $\frac{\langle Q^{2} \rangle}{V}$ = $(80 \pm 10)^{4}$ [MeV$^{4}$]. This result is reasonably consistent with our outcome. Therefore, the computations of the topological charges are adequately performed. In quenched QCD, the topological susceptibility is calculated and the outcome is $\frac{\langle Q^{2} \rangle}{V}$ = $(191 \pm 5)^{4}$ [MeV$^{4}$]~\cite{DeDebbio1}. This value is approximately 13 times larger than our outcome.

The instanton liquid model~\cite{Shuryak1_1} provides the instanton (or anti-instanton) density. Assuming CP invariance, we predict the number density of instantons and anti-instantons as $\frac{N_{I}}{V}$ = 1.6$\times10^{-3}$ [GeV$^{4}$]. From this prediction, we estimate that the number of instantons and anti-instantons in the physical lattice volume $V_{\text{phys}}$ = 29.9 [fm$^{4}$] ($V = 8^{3}\times16$) is $N_{I}^{\text{pred}}$ = 31.5. In quenched QCD, the agreement between the model and numerical calculations is confirmed~\cite{Hasegawa4,DiGH3}.

In this study, the number of instantons and anti-instantons of the standard configuration at low temperature is $N_{I}$ = 2.0(3) and its density is $\frac{N_{I}}{V}$ = 1.02(14)$\times10^{-4}$ [GeV$^{4}$]. These outcomes are approximately one-sixteenth of the values obtained from the quenched configurations or model calculations. Thus, the creation of zero modes is suppressed in full QCD, and the number of instantons and anti-instantons is much fewer than the results of quenched QCD or model calculations.

In quenched QCD, we have demonstrated that when the number density of instantons and anti-instantons and the low-mode density increase, chiral symmetry breaking is induced~\cite{Hasegawa2,Hasegawa4}. Therefore, the outcomes shown in Figs.~\ref{fig:Spec_dens_8xx3x16} and~\ref{fig:hist_Q} and Table~\ref{tb:top_ins} indicate that chiral symmetry breaking at low temperature would be emaciated by increasing the intensity of external magnetic fields. We will confirm this effect on chiral symmetry breaking in subsection~\ref{sec:chiral_symmetry_breaking}.

\subsection{Smearing effects on monopoles and spectra of the overlap Dirac operator at low temperature}\label{subsec:app2}

Smearing techniques are commonly used in generating configurations~\cite{Morningstar_1}, reducing statistical errors, and analyzing observables in lattice simulations~\cite{Allton_1,Y_Aoki3,Alexandrou_1,Teper_1}. The computational times of the overlap Dirac operator typically become very long; therefore, to smoothen fluctuations of configurations and reduce computational time, smearing is occasionally performed on the link variables a few times.
\begin{table*}[htbp]
  \begin{center}
    \caption{The numerical results at low temperature ($V = 8^{3}\times16$) are computed from the smeared link variables. The absolute value of the topological charges $\langle|Q|\rangle$, the number of instantons and anti-instantons $N_{I}$, the number density of instantons and anti-instantons $\frac{N_{I}}{V}$, and the topological susceptibility $\left(\frac{\langle Q^{2}\rangle}{V}\right)^{\frac{1}{4}}$.}\label{tb:app2_top}
    \begin{tabular}{|c|c|c|c|c|c|c|}\hline
      Smearing  & $(N, \alpha)$ & $\rho_{m}$ & $\langle|Q|\rangle$ & $N_{I}$ & $\frac{N_{I}}{V}$ & $\left(\frac{\langle Q^{2}\rangle}{V}\right)^{\frac{1}{4}}$ \\
       &  & [GeV$^{3}$] &  & &  [GeV$^{4}$] &  [MeV] \\ \hline       
      &   (6, 0.5)  & 3.99(2)$\times10^{-2}$ & 2.3(2) & 8.3(1.4) & 4.2(7)$\times10^{-4}$ & 143(6) \\\cline{2-7}
      &   (10, 0.5) & 3.94(2)$\times10^{-2}$ & 1.77(17) & 4.8(8) & 2.5(4)$\times10^{-4}$ & 125(5) \\ \cline{2-7}
      3D & (20, 0.5) & 3.93(2)$\times10^{-2}$ & 1.00(12) & 1.8(3)  & 9.3(17)$\times10^{-5}$ & 98(4) \\\cline{2-7}
      &   (30, 0.5) & 3.94(2)$\times10^{-2}$ & 0.75(9)  & 1.05(17) & 5.3(9)$\times10^{-5}$ & 85(4) \\ \cline{2-7}
      &   (40, 0.5) & 3.94(2)$\times10^{-2}$ & 0.45(8) & 0.58(13) & 3.0(7)$\times10^{-5}$ & 74(4) \\ \hline
      &   (6, 0.5)  & 3.58(2)$\times10^{-2}$ & 2.6(2) & 10.5(1.7) & 5.3(8)$\times10^{-4}$ &  152(6)   \\\cline{2-7}
      4D & (20, 0.5) & 2.97(2)$\times10^{-2}$ & 2.7(2) & 10.6(1.8) & 5.4(9)$\times10^{-4}$ & 152(7)  \\\cline{2-7}
      &   (40, 0.5) &  2.91(2)$\times10^{-2}$ & 2.5(2) &  9.8(1.7) & 5.0(9)$\times10^{-4}$ & 149(6)  \\ \hline
     \end{tabular}
  \end{center}
\end{table*}

One of the most frequently employed smearing techniques is APE~\cite{Ape1} smearing, which is used for determining the lattice spacing through the analysis of static potential. In subsection~\ref{sec:1_1}, we apply APE smearing to link variables and estimate the lattice spacing.

To clarify any unforeseen and unclear effects on observables that could arise from utilizing the smeared link variables, we apply APE smearing to link variables of standard configurations at low and high temperatures and prepare two types of conﬁgurations with the smeared link variables as described below: (1) smearing is applied to the spatial components of the lattice indicated as 3D. (2) Smearing is applied to the four-dimensional components of the lattice indicated as 4D.

We perform smearing to link variables varying the number of smearing repetitions $N$ from 6 to 40 and setting the weight factor to $\alpha = 0.5$, and calculate low-lying eigenvalues and eigenvectors of the massless overlap Dirac operator from these link variables.

We first compute the same observables as in the previous sections and then evaluate the chiral condensate, which is an order parameter for chiral symmetry breaking. Table~\ref{tb:app2_top} represents the numerical findings obtained using the configurations at low temperature.

After six rounds of the three-dimensional smearing, the monopole density decreases compared to the density of the configurations without smearing. However, the values remain unchanged when further smearing is performed. Additionally, the numerical results show that the absolute value of the topological charges, the number of instantons and anti-instantons, the number density of instantons and anti-instantons, and the topological susceptibility decrease with increasing number of iterations of the three-dimensional smearing.

In contrast with the numerical results of three-dimensional smearing, the monopole density decreases with increasing number of four-dimensional smearing iterations. However, the absolute value of the topological charges, the number density of instantons and anti-instantons, and the topological susceptibility remain unchanged.
\begin{table}[htbp]
  \begin{center}
    \caption{The fitting results of the distributions of the topological charges at low temperature ($V = 8^{3}\times16$) by the Gaussian function~(\ref{eq:gauss_1}). The configurations with the smeared link variables are used.}\label{tb:fit_top_2}
    \begin{tabular}{|c|c|c|c|c|}\hline
      Smearing  &  $(N, \alpha)$ & $\langle Q^{2}\rangle$ & $O(V^{-1})$ & $\chi^{2}/\text{dof}$ \\ \hline
      & (6, 0.5)  & 9(3) & -0.11(12) & 8/12  \\ \cline{2-5} 
      & (10, 0.5) & 5.3(1.4) &  -0.11(12) & 7/9    \\ \cline{2-5} 
      3D & (20, 0.5) & 1.8(5) & -0.03(13) & 2/5     \\ \cline{2-5} 
      & (30, 0.5) & 1.2(3) & 0.009(132) & 1/3 \\ \cline{2-5} 
      & (40, 0.5) & 0.40(10) & -0.06(13) & 3/3 \\ \hline
      & (6, 0.5)  & 14(4) & -0.007(141) & 6/12 \\ \cline{2-5} 
      4D & (20, 0.5) & 13(4) & -0.21(12)  & 14/14 \\ \cline{2-5} 
      & (40, 0.5) & 11(3) & -0.11(13) & 9/12 \\ \hline
    \end{tabular}
  \end{center}
\end{table}

We assume that the reason is that the monopole is a three-dimensional object, whereas the instanton is a four-dimensional object. Therefore, three-dimensional smearing does not affect monopoles, whereas it affects instantons. Four-dimensional smearing affects monopoles; however, it does not affect instantons\footnote{This idea comes from the discussion with E.-M. Ilgenfritz.}.

\begin{figure*}[htbp]
  \begin{center}
   \includegraphics[width=160mm]{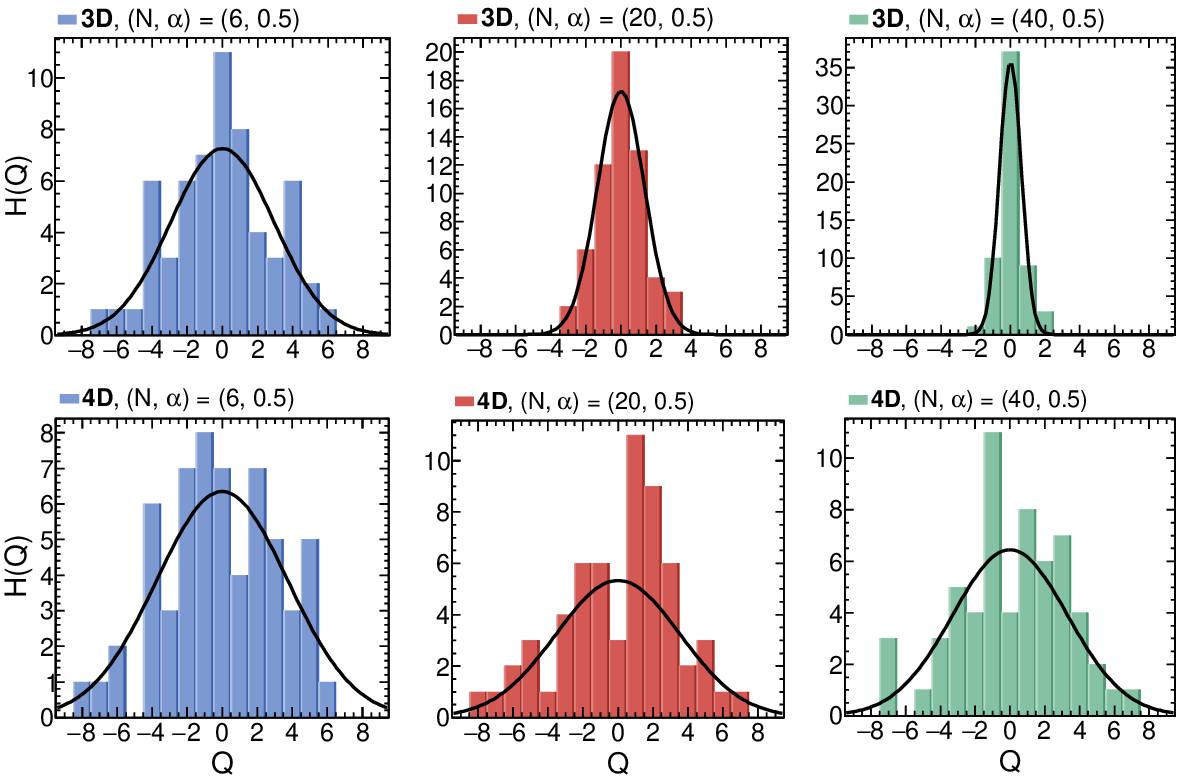}
  \end{center}
  \setlength\abovecaptionskip{-1pt}
  \caption{The topological charge distributions H($Q$) calculated from the configurations at low temperature ($V = 8^{3}\times16$) with the smeared link variables. The black curves indicate the results obtained by fitting the Gaussian function~(\ref{eq:gauss_1}).}\label{fig:H_Q_smear}
\end{figure*}

We find that the smearing technique affects the number of topological charges. Hence, we identify the topological charges present in the configurations with the smeared link variables and make the histograms of these charges. Table~\ref{tb:fit_top_2} presents the results obtained by fitting the Gaussian distribution function~(\ref{eq:gauss_1}) to the topological charge distributions at low temperature. The fitted outcomes and the observations in Fig.~\ref{fig:H_Q_smear} demonstrate that three-dimensional smearing significantly impacts the topological charge distributions, whereas the effects of four-dimensional smearing are comparatively minimal.
\begin{figure*}[htbp]
  \begin{center}
  \includegraphics[width=160mm]{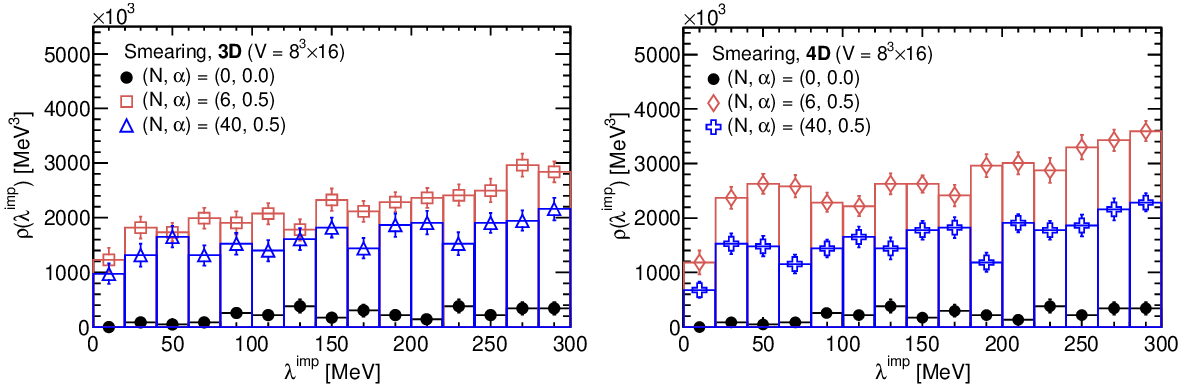}
  \end{center}
  \setlength\abovecaptionskip{-1pt}
  \caption{The spectral density $\rho(\lambda^{\text{imp}})$ at low temperature ($V = 8^{3}\times16$) compared to the outcomes calculated from the smeared link variables. The results of $(N, \alpha)$ = (0, 0.0) are the same results as those shown in Fig.~\ref{fig:Spec_dens_8xx3x16}, in which smearing is not applied.}\label{fig:spect_smear}
\end{figure*}

Next we investigate the impacts of smearing on the low-lying eigenvalues of the overlap Dirac operator at low temperature. We count the number of improved eigenvalues~(\ref{imp_op1}) at intervals of 20 [MeV] and calculate the spectral density~(\ref{eq:spec_1}). We compare the numerical findings with the outcomes of the configuration where no smearing was performed, as shown in Fig.~\ref{fig:spect_smear}. After six smearing iterations, the spectral densities reveal notable increases. However, the densities decrease subsequently after undergoing forty smearing iterations.

\begin{figure*}[htbp]
  \begin{center}
   \includegraphics[width=160mm]{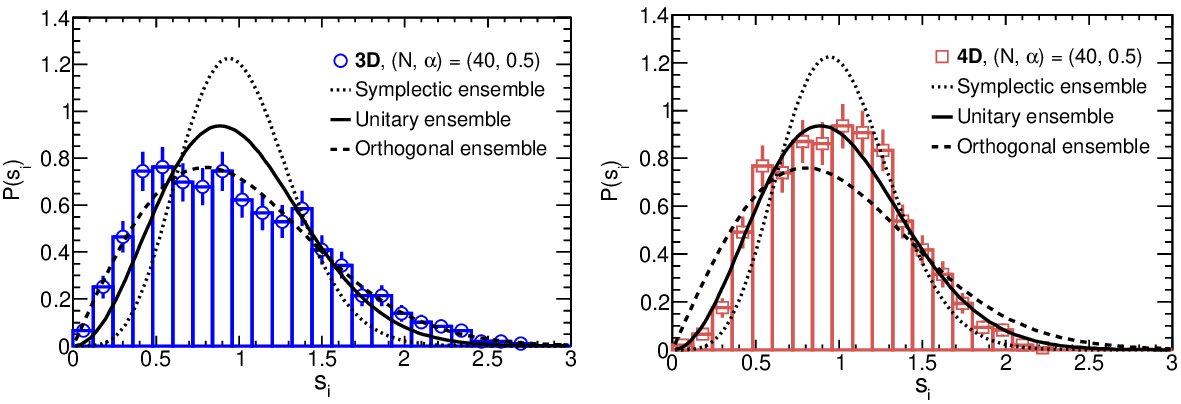}
  \end{center}
  \setlength\abovecaptionskip{-1pt}
  \caption{The distributions P($s_{i}$) of the nearest-neighbor spacing $s_{i}$ at low temperature ($V = 8^{3}\times16$) calculated using the smeared link variables. The parameters for smearing are $(N, \alpha) = (40, 0.5)$.}\label{fig:near_8xx3x16_sm_n40}
\end{figure*}

We then make the distributions of the nearest-neighbor spacing~(\ref{eq:near_si}) of the eigenvalues of the overlap Dirac operator calculated from the smeared link valuables. Figure~\ref{fig:near_8xx3x16_sm_n40} demonstrates that after forty rounds of three-dimensional smearing, the distribution of the nearest-neighbor spacing shifts from the distribution of the unitary ensemble to the distribution of the orthogonal ensemble, whereas after forty rounds of four-dimensional smearing, the distribution of the nearest-neighbor spacing is unchanged and its distribution agrees with the distribution of the unitary ensemble of GRMT.

We have demonstrated that smearing significantly impacts the monopoles and the spectra of the overlap Dirac operator. These monopoles, zero modes, and low-lying eigenvalues are closely related to chiral symmetry breaking~\cite{Hasegawa2,Hasegawa4}. Thus, we estimate the chiral condensate and examine the smearing effects on chiral symmetry breaking. In section~\ref{sec:sec4}, we will elucidate the computational details of the chiral condensate.

We calculate the chiral condensate~(\ref{eq:gmor_1}) from the Gell-Mann-Oakes-Renner (GMOR) relation~\cite{Gellmann1}, using approximately forty low-lying eigenvalues and eigenvectors for each configuration. In this study, the chiral condensate is defined as a negative value in the chiral limit. We vary the input mass of the quark propagator~(\ref{eq:quark_p}) within the range from 35 [MeV] to 100 [MeV] at intervals of 5 [MeV], calculate the chiral condensate, and extrapolate the numerical results in the chiral limit $am_{q} \rightarrow 0$ by fitting a linear function.

The numerical result of the configurations without smearing to the link variables is $\langle\bar{\psi}\psi\rangle$ = -1.44(9)$\times10^{-3}$ [GeV$^{3}$]. The numerical results of the configurations performing four-dimensional smearing are as follows: $(N, \alpha) = (6, 0.5)$, $\langle\bar{\psi}\psi\rangle$ = -7.2(3)$\times10^{-3}$ [GeV$^{3}$]. $(N, \alpha) = (40, 0.5)$, $\langle\bar{\psi}\psi\rangle$ = -4.8(2)$\times10^{-3}$ [GeV$^{3}$].

After applying smearing, the values of the chiral condensate in the chiral limit are three to five times greater than the value obtained without this process. Chiral symmetry breaking appears to be artificially induced by performing smearing to the link variables.

We carry out the same calculations as in this subsection using the configurations at high temperature and present the numerical results in section~\ref{sec:sm_finite}. The numerical outcomes demonstrate the significant effects on the observables caused by smearing to the link variables.

These results show that smearing affects the observables, and the number of iterations needed for measurements is uncertain. Therefore, except for examining the lattice spacing in subsection~\ref{sec:1_1}, we do not use any smearing techniques to calculate observables.

%%%%%%%%%%%%%%%%%%%%%%%%%%%%%%%%%%%%%%%%%%%%%%%%%%%%%%%%%%%%%%%%%%%%%%%%%%%
%% SEC 4
%%%%%%%%%%%%%%%%%%%%%%%%%%%%%%%%%%%%%%%%%%%%%%%%%%%%%%%%%%%%%%%%%%%%%%%%%%%

\section{Chiral symmetry breaking and eta-prime meson mass in external magnetic fields}\label{sec:sec4}

In this section, we calculate correlation functions for the pseudoscalar density using eigenvalues and eigenvectors at low temperature. We evaluate the impacts of external magnetic fields on the PCAC relation, decay constant, chiral condensate, and mass of the eta-prime meson.

\subsection{Correlation functions for the pseudoscalar density}

 In this research, we use the same notation of the overlap Dirac operator as in~\cite{Niedermayer1} and the same technique for calculating the correlation functions as in~\cite{Giusti2,DeGrand2,Wennekers1}. We compute the correlation functions for the pseudoscalar density similar to our previous research in quenched QCD~\cite{Hasegawa2,Hasegawa4,Hasegawa3}.

 The massive overlap Dirac operator $\mathcal{D}(m_{q})$, which has a valence quark mass $m_{q}$, is defined using the massless overlap Dirac operator~(\ref{eq:overlap_d}) as follows:
\begin{equation}
  \mathcal{D} (m_{q}) = \left(1 - \frac{am_{q}}{2\rho} \right)\mathcal{D} + m_{q}\label{eq:ovmass}
\end{equation}
The mass term $m_{q}$ is an input parameter. The massive eigenvalues $\lambda_{i}^{\text{mass}}$ of the massive overlap Dirac operator are calculated using the eigenvalues $\lambda_{i}$ of the massless overlap Dirac operator~(\ref{eq:overlap_d}) as follows:
\begin{equation}
\lambda_{i}^{\text{mass}} =\left(1-\frac{am_{q}}{2\rho} \right)\lambda_{i} + m_{q}\label{eq:mass_d}
\end{equation}

The quark propagator consists of low-lying eigenvalues and eigenvectors, as well as higher ones. The low modes dominate the meson correlation function~\cite{DeGrand2,Foley_1,Fukaya_1,Neff_1}; therefore, we use the following quark propagator derived from the representation of the eigenvalue decomposition~\cite{Foley_1}:
\begin{equation}
  G(\vec{y}, y^{0}; \vec{x}, x^{0}) \equiv \sum_{i}\frac{\psi_{i}(\vec{x}, x^{0}) \psi_{i}^{\dagger}(\vec{y}, y^{0})}{\lambda_{i}^{\text{mass}}}\label{eq:quark_p},
\end{equation}
where $\psi_{i}$ are the eigenvectors of the massless overlap Dirac operator. The subscript $i$ represents the number of eigenvalue and eigenvector pairs. When the numbers of zero modes are odd, we use 399 pairs. When the numbers of zero modes are even, we use 400 pairs.

The operator for the pseudoscalar density $\mathcal{O}_{PS}$ is defined as follows:
\begin{equation}
 \mathcal{O}_{PS}^{12} = \bar{\psi}_{1}\gamma_{5}\left( 1 - \frac{a}{2\rho}{\mathcal{D}} \right) \psi_{2}, \ \ \mathcal{O}_{PS}^{c, 21} = \bar{\psi}_{2}\gamma_{5}\left( 1 - \frac{a}{2\rho}\mathcal{D} \right) \psi_{1}.
\end{equation}
The superscript $c$ stands for the Hermitian adjoint of the operator. For convenience, indices 1 and 2 indicate quark ﬂavors~\footnote{The notation for the massive overlap Dirac operator~(\ref{eq:ovmass}) does not have quark flavors.}.

We define the correlation function $C(\Delta t)_{PS}$ for the connected contribution as follows:
\begin{equation}
  C(\Delta t)_{PS} = \frac{a^{3}}{V}  \sum_{t}\sum_{\vec{x}_{2}}\sum_{\vec{x}_{1}}\left\langle\mathcal{O}_{PS}^{c, 21}(\vec{x}_{2}, t) \mathcal{O}_{PS}^{12}(\vec{x}_{1}, t + \Delta t)\right\rangle\label{eq:corre_conne}.
\end{equation}
Similar to the connected contribution, we define the correlation function $C_{DPS}(\Delta t)$ for the disconnected contribution as follows:
\begin{equation}
  C_{DPS}(\Delta t) = \frac{a^{3}}{V} \sum_{t}\sum_{\vec{x}_{2}}\sum_{\vec{x}_{1}}\left\langle\mathcal{O}_{PS}^{11}(\vec{x}_{2}, t)\mathcal{O}_{PS}^{22}(\vec{x}_{1}, t + \Delta t)\right\rangle\label{eq:corre_disconne}
\end{equation}
To reduce statistical errors~\cite{DeGrand2,Foley_1}, we calculate correlations among all components of the correlation functions of $C_{PS}(\Delta t)$ and $C_{DPS}(\Delta t)$, and sum the spatial components $\vec{x}_{1}$, $\vec{x}_{2}$ and a temporal component $t$, and estimate statistical errors by using the Jackknife method.

\subsection{PCAC relation and decay constant}

 In this study, the statistical errors are very large because the statistics are not sufficient; therefore, we calculate the correlation functions~(\ref{eq:corre_conne}) and~(\ref{eq:corre_disconne}) varying the input quark mass $m_{q}$ of formula~(\ref{eq:mass_d}) from 35 [MeV] to 100 [MeV] at intervals of 1 [MeV], and extrapolate 66 data points of the numerical results in the chiral limit to reduce statistical errors. The bare quark masses of the configurations are fixed.
\begin{table*}[htbp]
  \begin{center}
    \caption{The fitting results of the PCAC relation. $FR$ indicates the fitting range.}\label{tb:fit_res_pcac}
    \begin{tabular}{|c|c|c|c|c|c|}\hline
      $V$ & $Bz$ & $aA$ & $a^{2}B$ & $FR:$ [$am_{q}$] & $\chi^{2}/\text{dof}$ \\ \hline
          &  0 & 1.193(15) & -2.0(13)$\times10^{-3}$ & 0.052-0.126 & 61.0/57.0 \\ \cline{2-6}
      8$^{3}\times$16 & 3 & 1.13(2) & 6(17)$\times10^{-4}$ & 0.041-0.109 &  41.0/51.0     \\ \cline{2-6}
                   &   6 &  1.22(3) & 3(26)$\times10^{-4}$ & 0.041-0.104 & 35.5/47.0      \\ \hline
    \end{tabular}
  \end{center}
\end{table*}
\begin{figure*}[htbp]
  \begin{center}
    \includegraphics[width=150mm]{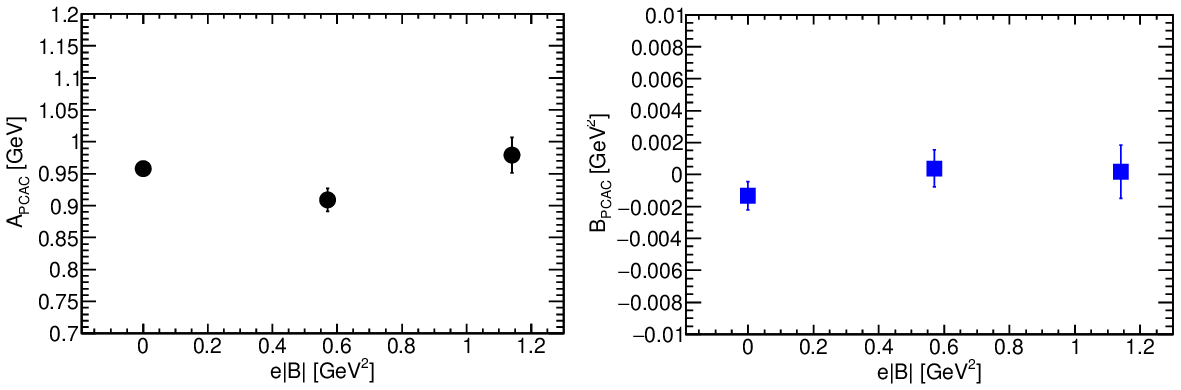}
  \end{center}
  \setlength\abovecaptionskip{-1pt}
  \caption{The fitting results of the slope $A_{\text{PCAC}}$ (left) and intercept $B_{\text{PCAC}}$ (right) of the PCAC relation.}\label{fig:pcac}
\end{figure*}

We assume that the correlation function~(\ref{eq:corre_conne}) of the connected contribution can be approximated using the following curve:
\begin{equation}
  C_{PS}(t) = \frac{a^{4}Z_{PS}}{am_{PS}} \exp\left( {-\frac{m_{PS}}{2}T}\right) \cosh\left[m_{PS} \left(\frac{T}{2} - t \right) \right]\label{eq:fit_func_cone}.
\end{equation}
We obtain the fitting parameters $a^{4}Z_{PS}$ and $am_{PS}$ by fitting this curve to the numerical results. The fitting results of $a^{4}Z_{PS}$ and $am_{PS}$ are represented in Tables 2-7 of~\cite{Hasegawa7}.

First, we confirm the impact of external magnetic fields on the PCAC relation by fitting the following function to the numerical results:
\begin{equation}
  a^{2}m_{PS}^{2} = aA_{\text{PCAC}}x + a^{2}B_{\text{PCAC}}, \ x = am_{q}.
\end{equation}
The fitting range is determined so that the value of $\chi^{2}/\text{dof}$ is approximately 1. The fitting results are listed in Table~\ref{tb:fit_res_pcac}. The left panel of Fig.~\ref{fig:pcac} shows that the slope values $A_{\text{PCAC}}$ of the PCAC relation are not affected by external magnetic fields. The right panel of Fig.~\ref{fig:pcac} shows that the fitting results of the intercept $B_{\text{PCAC}}$ are approximately zero regarding their errors and external magnetic fields do not affect the intercept $B_{\text{PCAC}}$.
\begin{table*}[htbp]
  \begin{center}
    \caption{The fitting results of the decay constant $F_{0}$ in the chiral limit of the pseudoscalar.}\label{tb:res_fps}
    \begin{tabular}{|c|c|c|c|c|c|c|}\hline
      $V$ & $Bz$ & $a^{-1}A_{F_{PS}}$ & $aB_{F_{PS}}$ & $F_{0}$ & $FR:$ & $\chi^{2}/\text{dof}$ \\
       & & $\times10^{-2}$ & $\times10^{-2}$ & [MeV] & [$(am_{PS})^{2}$] & \\ \hline
           &  0 & 5.3(14) & 6.64(15) & 53.3(1.2) & 0.056-0.146 & 4.6/64.0   \\ \cline{2-7}
      8$^{3}\times$16 & 3 & 5.1(16) & 6.47(18) & 52.0(1.5) & 0.054-0.142 & 3.6/64.0   \\ \cline{2-7}
                    &   6 & 4.1(1.7) & 5.9(2) & 47.7(1.6) & 0.057-0.154 & 3.4/64.0   \\ \hline
    \end{tabular}
  \end{center}
\end{table*}
\begin{figure*}[htbp]
  \begin{center}
    \includegraphics[width=73mm]{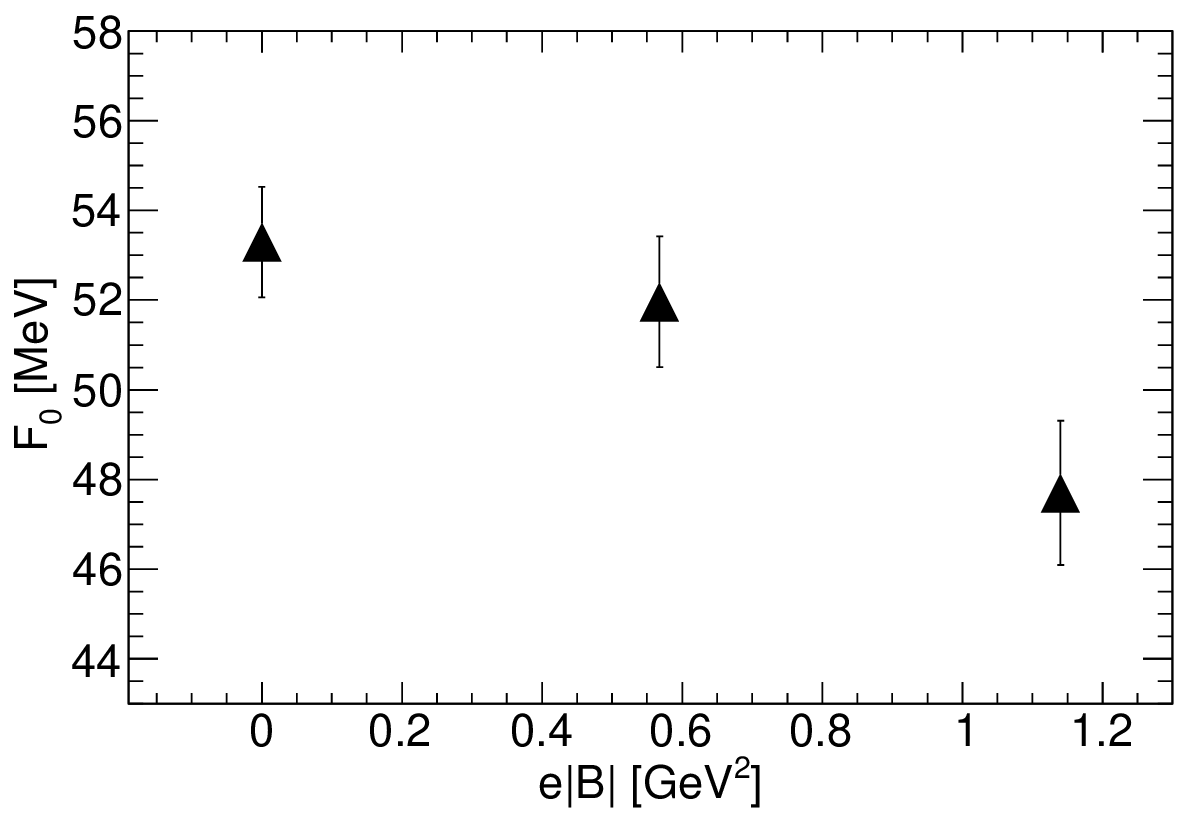}
  \end{center}
  \setlength\abovecaptionskip{-1pt}
  \caption{The decay constant $F_{0}$  in the chiral limit of the pseudoscalar.}\label{fig:fpi2}
\end{figure*}

Then, we evaluate the impact of external magnetic fields on the decay constant of the pseudoscalar, which is defined as follows:
\begin{equation}
aF_{PS} = \frac{2am_{q}\sqrt{a^{4}G_{PS-SS}}}{(am_{PS})^{2}}\label{eq:fps}
\end{equation}
In this notation, the pion decay constant is $F_{\pi}$ = 93 [MeV]. The calculation results of the decay constant $aF_{PS}$ are listed in Tables 2-7 in~\cite{Hasegawa7}.

To assess the effect, we fit the following function to the numerical outcomes and obtain the extrapolated outcomes in the chiral limit:
\begin{equation}
  aF_{PS} = \frac{A_{F_{PS}}}{a}x + aB_{F_{PS}}, \ x = (am_{PS})^{2}.
\end{equation}
The fitting ranges include all data points and the outcomes are displayed in Table~\ref{tb:res_fps}. Figure~\ref{fig:fpi2} demonstrates that the decay constant $F_{0}$ in the chiral limit of the pseudoscalar decreases as the intensity of external magnetic fields increases.

\subsection{Chiral symmetry breaking}\label{sec:chiral_symmetry_breaking}

Next we calculate the chiral condensate that is derived from the Gell-Mann-Oakes-Renner (GMOR) relation~\cite{Gellmann1} as follows:
\begin{equation}
  a^{3}\langle{\bar\psi} \psi \rangle = - \lim_{(am_{PS})^{2} \rightarrow 0}\frac{(am_{PS}aF_{PS})^{2}}{2am_{q}}\label{eq:gmor_1}
\end{equation}
In this definition, the chiral condensate is negative and defined in the chiral limit when $(am_{PS})^{2}\rightarrow 0$ instead of $am_{q}\rightarrow 0$. This is due to the confirmed linear increase in the PCAC relation.
\begin{table*}[htbp]
  \begin{center}
    \caption{The fitting results of the chiral condensate in the chiral limit and the dimensionless ratio $R(Bz)$. The estimated errors of the dimensionless ratio are calculated from the relative errors of the fitting results of $a^{3}B_{\langle\bar{\psi}\psi\rangle}$.}\label{tb:fit_res_chiral}
    \begin{tabular}{|c|c|c|c|c|c|c|c|}\hline
      $V$ & $Bz$ & $aA_{\langle\bar{\psi}\psi\rangle}$ & $a^{3}B_{\langle\bar{\psi}\psi\rangle}$ & $\langle\bar{\psi}\psi\rangle$ & $R(B_{z})$ & $FR:$  & $\chi^{2}/\text{dof}$ \\
      &  &  & &  [GeV$^{3}$] &  & [$(am_{PS})^{2}$] & \\
      &  & $\times10^{-3}$ & $\times10^{-2}$ & $\times10^{-3}$ & &  & \\ \hline
      & 0 &  4.8(6) & -2.52(6) & -1.31(3) & 0.00(0) & 0.057-0.146 & 39.0/64.0 \\ \cline{2-8}
      8$^{3}\times$16 & 3 & 4.7(7) & -2.30(7) & -1.19(4) & 8.8(4)$\times10^{-2}$ & 0.054-0.142 & 25.3/64.0 \\ \cline{2-8}
                    &   6 & 4.3(7) & -2.03(8) & -1.05(4) &  0.195(9) & 0.057-0.154 & 19.7/64.0 \\ \hline
    \end{tabular}
  \end{center}
\end{table*}
\begin{figure*}[htbp]
  \begin{center}
    \includegraphics[width=155mm]{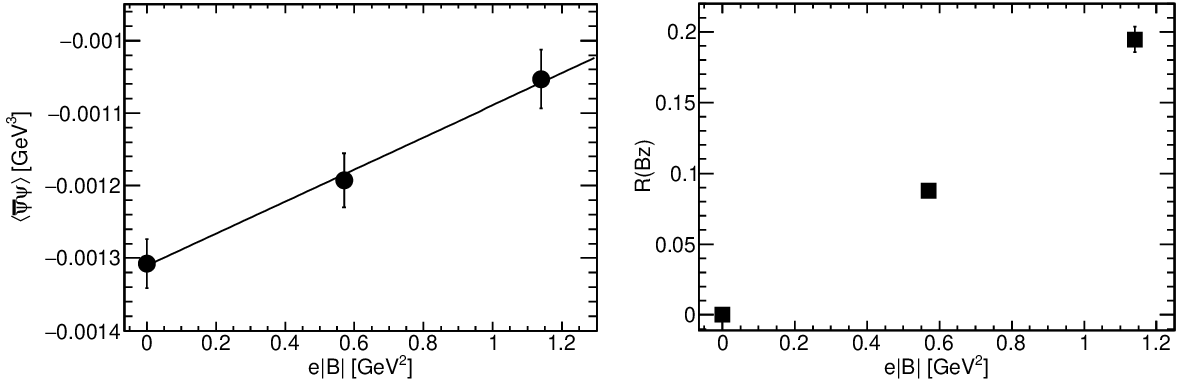}
  \end{center}
  \setlength\abovecaptionskip{-1pt}
  \caption{The chiral condensate $\langle\bar{\psi}\psi\rangle$ in the chiral limit (left) and the dimensionless ratio $R(Bz)$ of the chiral condensate (right).}\label{fig:chiral_condensate}
\end{figure*}

The calculation results of the chiral condensate $a^{3}\langle{\bar\psi} \psi \rangle$ are listed in Tables 2-7 in~\cite{Hasegawa7}. We fit the following curve to the numerical results and obtain the extrapolated results in the chiral limit:
\begin{equation}
  a^{3}\langle\bar{\psi}\psi\rangle = aA_{\langle\bar{\psi}\psi\rangle}x + a^{3}B_{\langle\bar{\psi}\psi\rangle}, \ x = (am_{PS})^{2}.
\end{equation}
All data points are included within the fitting ranges and the fitting results are presented in Table~\ref{tb:fit_res_chiral}.

To quantify the increases, we calculate the dimensionless ratio $R(Bz)$ of the chiral condensate in the chiral limit, which is defined as
\begin{equation}
  R(Bz) = \frac{B_{\langle\bar{\psi}\psi\rangle Bz} - B_{\langle\bar{\psi}\psi\rangle Bz = 0}}{B_{\langle\bar{\psi}\psi\rangle Bz =0}}\label{eq:ratio}.
\end{equation}
$B_{\langle\bar{\psi}\psi\rangle Bz}$ represents the fitting outcomes for external magnetic fields of $Bz$, as presented in Table~\ref{tb:fit_res_chiral}. The left panel of Fig.~\ref{fig:chiral_condensate} shows that the chiral condensate $\langle\bar{\psi}\psi\rangle$ in the chiral limit increases with the intensity of external magnetic fields. The right panel of Fig.~\ref{fig:chiral_condensate} demonstrates that the dimensionless ratio $R(Bz)$ of the chiral condensate in the chiral limit increases with the intensity of external magnetic fields.

The chiral condensate is defined as a negative value. There is no theoretical basis for the linear increase in the chiral condensate in the chiral limit in this study. However, we assume the linearity and fit the following linear function to the numerical outcomes: $y = A_{R}x - B_{R}$, ($x = e|B|$ [GeV$^{2}$]), as depicted in the left panel of Fig~\ref{fig:chiral_condensate}. The fitting results, $A_{R} = 2.2(5)\times10^{-4}$ [GeV], $B_{R} = 1.31(3)\times10^{-3}$ [GeV$^{3}$], and $\chi^{2}/\text{dof} = 0.1/1.0$, suggest that the chiral condensate in the chiral limit approaches zero at $e|B| \sim 6$ [GeV$^{2}$].

\subsection{Eta-prime meson}

Finally, we estimate the mass of the eta-prime meson by two methods. First, the mass of the eta-prime meson is estimated using the Witten-Veneziano mass formula~\cite{Giusti8}, the numerical results of the topological susceptibility $\frac{\langle Q^{2}\rangle}{V}$ provided in Table~\ref{tb:top_ins}, and the numerical results of $F_{0}$ provided in Table~\ref{tb:res_fps} as follows:\footnote{Here, we use the numerical result of $F_{0}$ instead of $F_{\pi}$ because the statistical errors of the pion decay constant $F_{\pi}$ are very large.} 
\begin{equation}
m_{\eta'}^{2} = \frac{2N_{f}}{F_{0}^{2}}\frac{\langle Q^{2}\rangle}{V}
\end{equation}
The computation results of $m_{\eta'}^{2}$ with $N_{f} = 3$ are in Table~\ref{tb:fit_res_eta}. The computational errors of $m_{\eta'}^{2}$ are from 14$\%$ to 18$\%$; therefore, we do not observe the impact of external magnetic fields on the square mass of the eta-prime meson.

Second, along the lines of the studies on estimating the mass of the eta-prime meson in quenched and full QCD~\cite{DeGrand3,Hashimoto_1}, we estimate the mass of the ﬂavor singlet pseudoscalar meson $\eta'$ from the correlation functions of~(\ref{eq:corre_conne}) for the connected contribution and~(\ref{eq:corre_disconne}) for the disconnected contribution.
\begin{table*}[htbp]
  \begin{center}
    \caption{The square mass $m_{\eta'}^{2}$ of the eta-prime meson and the fitting results for the square mass $\mu_{0}^{2}$ of the eta-prime meson.}\label{tb:fit_res_eta}
    \begin{tabular}{|c|c|c|c|c|c|c|c|}\hline
      $V$ & $Bz$ & $m_{\eta'}^{2}$ & $A_{\eta'}$ & $a^{2}B_{\eta'}$ & $\mu_{0}^{2}$ & $FR:$ & $\chi^{2}/\text{dof}$ \\
          &      & [GeV$^{2}$]    &            & $\times10^{-2}$ & [GeV$^{2}$]  & [$(am_{PS})^{2}$] &  \\ \hline
                     &  0 & 0.21(3) & -0.45(6) & 8.2(7) & 0.213(18) & 0.056-0.146 & 6.6/64.0  \\ \cline{2-8}
      8$^{3}\times$16 & 3 & 0.22(4) & -0.41(7) & 7.1(8) & 0.18(2)   & 0.054-0.142 & 3.4/64.0    \\ \cline{2-8}
                     & 6 & 0.23(4) &  -0.37(6) & 7.0(8) & 0.18(2)   & 0.057-0.154 & 3.1/64.0  \\ \hline
    \end{tabular}
  \end{center}
\end{table*}

The correlation function of the eta-prime meson consists of two parts: the connected contribution $C(\Delta t)$ and the disconnected contribution $D(\Delta t)$ for the pseudoscalar density~\footnote{The notation of refs~\cite{DeGrand3,Hashimoto_1} is followed.}.
\begin{equation}
  \sum_{\vec{x}_{1}}\sum_{\vec{x}_{2}} \sum_{t}\langle\eta(\vec{x}_{2}, t) \eta^{\dagger}(\vec{x}_{1}, t+\Delta t) \rangle = C(\Delta t) - N_{f}D(\Delta t)\label{eq:corre_eta_1}
\end{equation}
In quenched QCD, this correlation function in the momentum space $p$ with a double pole is expressed as follows:
\begin{equation}
  \langle\eta(p) \eta^{\dagger}(p) \rangle \propto \frac{1}{p^{2} + m_{PS}^{2}} - N_{f}\frac{1}{p^{2} + m_{PS}^{2}}\frac{\mu_{0}^{2}}{N_{f}}\frac{1}{p^{2} + m_{PS}^{2}}\label{eq:corre_eta}
\end{equation}
$\mu_{0}^{2}$ is the square mass of the eta-prime meson in quenched QCD. The first and second terms of the right-hand side of equation~(\ref{eq:corre_eta}) come from the connected and disconnected contributions, respectively. The second term is derived by differentiating the first term as follows:
\begin{equation}
  D(p) = -\frac{\mu_{0}^{2}}{2m}\frac{\partial}{\partial m}C(p)~\label{eq:calcu1}
\end{equation}

We have verified that the curve~(\ref{eq:fit_func_cone}) provides an approximation of the correlation function~(\ref{eq:corre_conne}) of the connected contribution. Consequently, we derive the following curve by differentiating the curve~(\ref{eq:fit_func_cone}) to fit the correlation function~(\ref{eq:corre_disconne}) of the disconnected contribution:
\begin{align}
  & C_{DPS}(t) = \frac{a^{6}Z_{DPS}}{(am_{PS_{2}})^{3}}\left\{(1 + m_{PS_{2}}t)\exp(-m_{PS_{2}}t) + \left[1 + m_{PS_{2}}(T - t)\right]\exp\left[-m_{PS_{2}}(T - t)\right]\right\}\label{eq:fit_func_discone},\\
  & a^{6}Z_{DPS} = \frac{a^{4}Z_{PS}(a\mu_{0})^{2}}{4}\label{eq:fit_func_z2}.
\end{align}

We fit the curve~(\ref{eq:fit_func_discone}) to the numerical results and acquire the fitting results of $a^{6}Z_{DPS}$ and $am_{PS_{2}}$. We then substitute the fitting results of $a^{4}Z_{PS}$ (which are represented in Tables 2-7 of~\cite{Hasegawa7}) and $a^{6}Z_{DPS}$ into equation~(\ref{eq:fit_func_z2}) and calculate $\frac{(a\mu_{0})^{2}}{4}$. The fitting results of $a^{6}Z_{DPS}$ and calculation results of $\frac{(a\mu_{0})^{2}}{4}$ are listed in Tables 8-10 in~\cite{Hasegawa7}.

We extrapolate the calculated results of $\frac{(a\mu_{0})^{2}}{4}$ in the chiral limit by the following linear function:
\begin{equation}
  \frac{(a\mu_{0})^{2}}{4} = A_{\eta'}x + a^{2}B_{\eta'}, \ x = (am_{ps})^{2}.
\end{equation}
\begin{figure*}[htbp]
  \begin{center}
    \includegraphics[width=75mm]{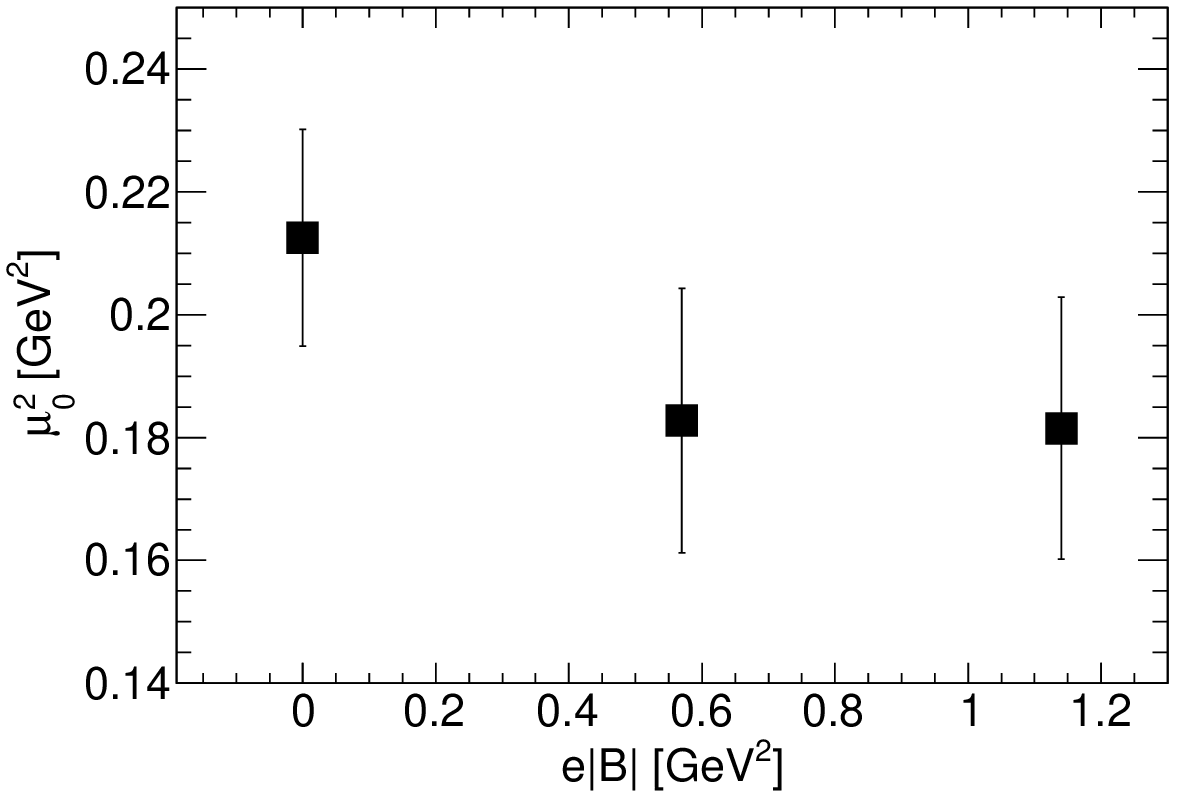}
  \end{center}
  \setlength\abovecaptionskip{-1pt}
  \caption{The effects of external magnetic fields on the square mass $\mu_{0}^{2}$ in the chiral limit of the eta-prime meson.}\label{fig:eta-prime}
\end{figure*}

All calculated results are included within the ﬁtting ranges and Table~\ref{tb:fit_res_eta} displays the fitting results. The extrapolated result of the square mass $\mu_{0}^{2}$ in the chiral limit of the normal configuration corresponds to the outcomes of $m_{\eta'}^{2}$ represented in Table~\ref{tb:fit_res_eta}.  Figure~\ref{fig:eta-prime} demonstrates that the square mass $\mu_{0}^{2}$ in the chiral limit of the eta-prime meson decreases slightly as the intensity of external magnetic fields increases.

%%%%%%%%%%%%%%%%%%%%%%%%%%%%%%%%%%%%%%%%%%%%%%%%%%%%%%%%%%%%%%%%%%%%%%%%%%%
%% SEC 5
%%%%%%%%%%%%%%%%%%%%%%%%%%%%%%%%%%%%%%%%%%%%%%%%%%%%%%%%%%%%%%%%%%%%%%%%%%%

\section{Summary and conclusion}\label{sec:sec5}

We have performed numerical calculations that were intended for initial computations using lattices of small volumes of coarse lattice spacing under the research collaboration. We investigated the effects of external magnetic fields on the monopoles, spectra of the overlap Dirac operator, instantons, chiral symmetry breaking, and the mass of the eta-prime meson.

First, we examined the effect of external magnetic fields on the longest monopole loops at low and high temperatures. Our findings revealed that the length of the longest monopole loops increases at low temperature as the intensity of external magnetic fields increases while it decreases at high temperature. Moreover, the absolute value of the Polyakov loops at low temperature remains unchanged when the strength of external magnetic fields grows while the value rises at high temperature.

Second, we calculated the eigenvalues and eigenvectors of the overlap Dirac operator. We demonstrated that the spectral density at low temperature slightly decreases with increasing the intensity of external magnetic fields while the spectral density at high temperature is slightly affected only around the gap near $\lambda^{\text{imp}}$ $\sim$ 900 [MeV]. We demonstrated that in full QCD at high temperature, the spectral density near zero eigenvalues is suppressed compared to the spectral density in quenched QCD.

Next we examined the impacts of external magnetic fields on eigenvalue fluctuations by comparing the distributions of the nearest-neighbor spacing and the computed spectral rigidity results with the GRMT predictions. Our study demonstrated that the distributions of the nearest-neighbor spacing at both low and high temperatures remain consistent with the GRMT prediction, indicating that external magnetic fields do not influence them. We found that, at low temperature, the spectral rigidity is consistent with the GRMT prediction. However, at high temperature, the spectral rigidity does not agree with the GRMT predictions and external magnetic fields affect the spectral rigidity.

Overlap fermions have exact zero modes in their spectra. Therefore, we inspected the influence of external magnetic fields on the topological charges and instantons. We made histograms of the topological charges and estimated the number of instantons and anti-instantons from the topological charge distributions. We did not detect a significant impact of external magnetic fields on the number of instantons and anti-instantons because the statistics are insufficient.

However, we discovered that the chiral magnetic effect at low temperature is that the formation of the topological charges changes because the number of zero modes decreases slightly when the intensity of external magnetic fields increases. As a result, the number of instantons and anti-instantons decreases at low temperature with increasing strength of external magnetic fields. At low temperature, we demonstrated that the spectral density and the number of instantons and anti-instantons decline as the intensity of external magnetic fields increases. This result suggests that intensifying magnetic fields weakens chiral symmetry breaking. In addition, we demonstrated that in full QCD, the creations of zero modes are suppressed compared to the outcomes of quenched QCD and model calculations.

We demonstrated that the number of instantons and anti-instantons increases with lengthening the total physical length of monopole loops in the study of quenched QCD~\cite{DiGH4}. However, in this study, we showed that at low temperature the monopole density increases while the number of instantons and anti-instantons decreases with increasing the intensity of external magnetic fields. This finding conflicts with prior research, and we suppose that this difference comes from the effects of magnetic fields on monopoles and instantons.

Next to clarify any unforeseen and unclear effects on observables that could arise from using the smeared link variables, we performed APE smearing to the link variables of standard configurations at low and high temperatures. We demonstrated significant effects on monopoles, spectra of the overlap Dirac operator, and chiral condensate in the chiral limit caused by smearing to link variables.

Third, we calculated the correlation functions of the connected and disconnected contributions for the pseudoscalar density at low temperature. We evaluated the impact of external magnetic fields on the PCAC relation and showed that the PCAC relation is not affected by external magnetic fields. Next we evaluated the decay constant in the chiral limit of the pseudoscalar meson and demonstrated that it decreases with increasing the intensity of external magnetic fields.

We then computed the chiral condensate, extrapolated the outcomes in the chiral limit, and calculated the dimensionless ratio of the chiral condensate in the chiral limit using the extrapolation results. We found that the chiral condensate in the chiral limit and the dimensionless ratio increase when the intensity of external magnetic fields increases. The chiral condensate in the chiral limit is a negative value. Therefore, it is assumed that the chiral condensate in the chiral limit becomes zero when the intensity of external magnetic fields reaches approximately $e|B| \sim 6$ [GeV$^{2}$].

Finally, we estimated the square mass of the eta-prime meson using two methods. The first method is its estimation using the Witten-Veneziano mass formula and the numerical results of the decay constant in the chiral limit of the pseudoscalar meson and the topological susceptibility. In the second method, we estimated the square mass from the correlation functions of the connected and disconnected contributions. Moreover, we reduced the statistical errors by extrapolating the outcomes computed from 66 bare quark masses in the chiral limit.

The outcomes of the square mass of the eta-prime meson using the standard configurations agree between both methods. However, in the first method, we showed that the statistical errors are larger than 14$\%$; therefore, we could not demonstrate the influence of external magnetic fields on the mass of the eta-prime meson. 

In the second method, we demonstrated that in the chiral limit the square mass of the eta-prime meson becomes slightly lighter with increasing the intensity of external magnetic fields.

However, the conclusions of this study are affected by the finite lattice volume and discretization. Statistical errors remain significant. Therefore, it is imperative to refine our computations using the lattices of larger lattice volumes of a finer lattice spacing with stronger magnetic fields than those in this study and increasing statistics.

%%%%%%%%%%%%%%%%%%%%%%%%%%%%%%%%%%%%%%%%%%%%%%%%%%%%%%%%%%%%%%%%%%%%%%%%%%%
%% SEC Appendix
%%%%%%%%%%%%%%%%%%%%%%%%%%%%%%%%%%%%%%%%%%%%%%%%%%%%%%%%%%%%%%%%%%%%%%%%%%%

\appendix

\section{Smearing effects on observables at high temperature}\label{sec:sm_finite}

This section presents the numerical results obtained from the configurations at high temperature ($V = 8^{3}\times4$) with the smeared link variables. We compute the same observables explained in subsection~\ref{subsec:app2}. We exhibit the numerical results, Tables, and Figures in this section.
\begin{table*}[htbp]
  \begin{center}
    \caption{Identical measurements as those shown in Table~\ref{tb:app2_top}. These results are computed using the configurations at high temperature ($V = 8^{3}\times4$) with the smeared link variables.}\label{tb:app2_top_fin}
    \begin{tabular}{|c|c|c|c|c|c|c|}\hline
      Smearing  & $(N, \alpha)$ & $\rho_{m}$ & $\langle|Q|\rangle$ & $N_{I}$ & $\frac{N_{I}}{V}$ & $\left(\frac{\langle Q^{2}\rangle}{V}\right)^{\frac{1}{4}}$ \\
       &  & [GeV$^{3}$] &  & &  [GeV$^{4}$] &  [MeV] \\ \hline       
      &    (6, 0.5)  & 2.94(5)$\times10^{-2}$ & 0.87(9)   &  1.2(2)   & 2.4(5)$\times10^{-4}$   & 125(6)  \\\cline{2-7}
      3D & (20, 0.5) & 2.89(5)$\times10^{-2}$ & 0.53(8)   &  0.67(13) & 1.4(3)$\times10^{-4}$   & 108(5)  \\\cline{2-7}
      &    (40, 0.5) & 2.87(5)$\times10^{-2}$ & 0.25(6)   &  0.28(8)  & 5.8(1.7)$\times10^{-5}$ & 87(6)  \\ \hline
      &    (6, 0.5)  & 2.55(4)$\times10^{-2}$ & 0.63(10)  &  1.0(2)   & 2.1(5)$\times10^{-4}$   & 120(7)  \\\cline{2-7}
      4D & (20, 0.5) & 1.95(4)$\times10^{-2}$ & 0.37(7)   &  0.43(11) & 9(2)$\times10^{-5}$     & 97(6)   \\\cline{2-7}
      &    (40, 0.5) & 1.86(4)$\times10^{-2}$ & 0.23(6)   &  0.27(8)  & 5.4(1.7)$\times10^{-5}$ & 86(7)   \\ \hline
     \end{tabular}
  \end{center}
\end{table*}   
\begin{figure*}[htbp]
  \begin{center}
   \includegraphics[width=157mm]{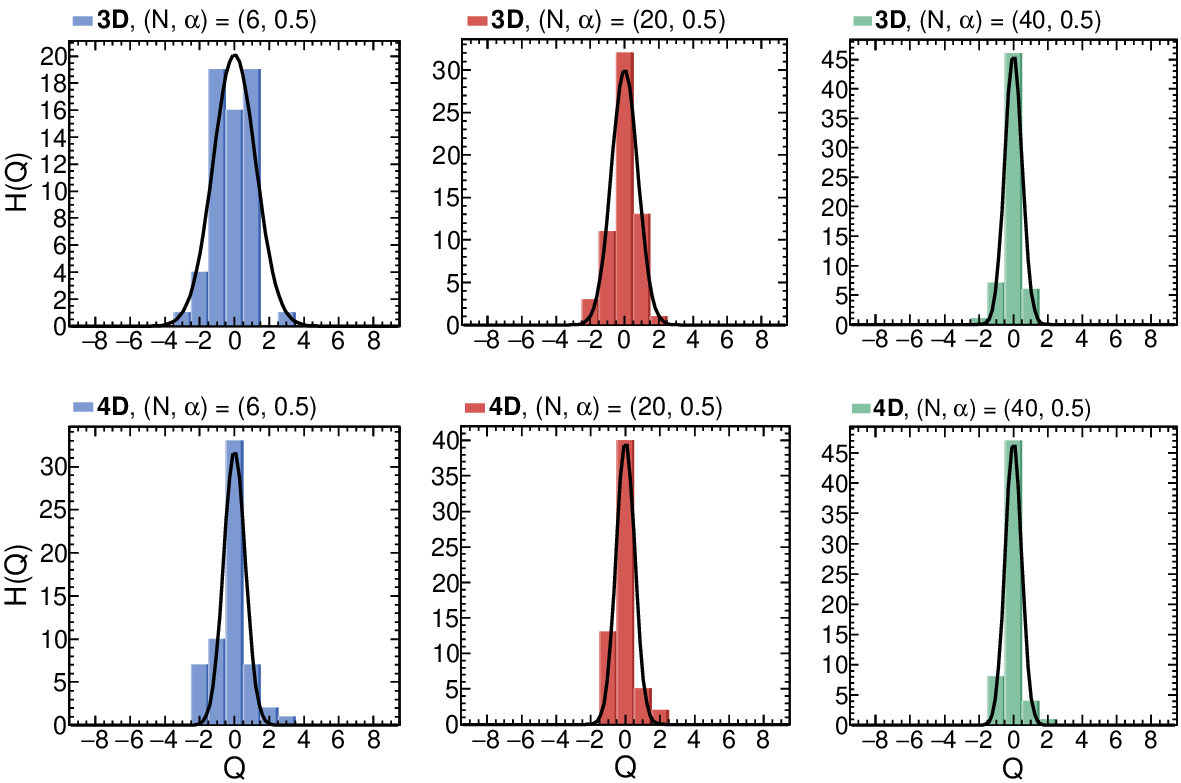}
  \end{center}
  \setlength\abovecaptionskip{-1pt}
  \caption{The topological charge distributions H($Q$) at high temperature ($V = 8^{3}\times4$). The topological charges $Q$ are calculated from the configurations with the smeared link variables. The black curves show the outcomes acquired by fitting the Gaussian function~(\ref{eq:gauss_1}). Table~\ref{tb:fit_top_3} exhibits the fitted results.}\label{fig:H_Q_smear_2}
\end{figure*}
\begin{table}[htbp]
  \begin{center}
    \caption{The fitting results of the distributions H($Q$) of the topological charges $Q$ at high temperature ($V = 8^{3}\times4$), as shown in Fig.~\ref{fig:H_Q_smear_2}, obtained by fitting the Gaussian function~(\ref{eq:gauss_1}). The configurations with the smeared link variables are used.}\label{tb:fit_top_3}
    \begin{tabular}{|c|c|c|c|c|}\hline
      Smearing  &  $(N, \alpha)$ & $\langle Q^{2}\rangle$ & $O(V^{-1})$ & $\chi^{2}/\text{dof}$ \\ \hline
      & (6, 0.5)  & 1.5(3)  &  0.03(14) & 4/4  \\ \cline{2-5} 
  3D & (20, 0.5)  & 0.60(14)& -0.03(13) & 2/3  \\ \cline{2-5} 
      & (40, 0.5) & 0.26(4) & -0.03(13) & 1/2  \\ \hline
      & (6, 0.5)  & 0.39(11)& -0.17(12) & 10/4 \\ \cline{2-5} 
   4D & (20, 0.5) & 0.29(5) & -0.10(12) & 5/2  \\ \cline{2-5} 
      & (40, 0.5) & 0.23(4) & -0.06(12) & 2/2  \\ \hline
    \end{tabular}
  \end{center}
\end{table}
\begin{figure*}[htbp]
  \begin{center}
  \includegraphics[width=160mm]{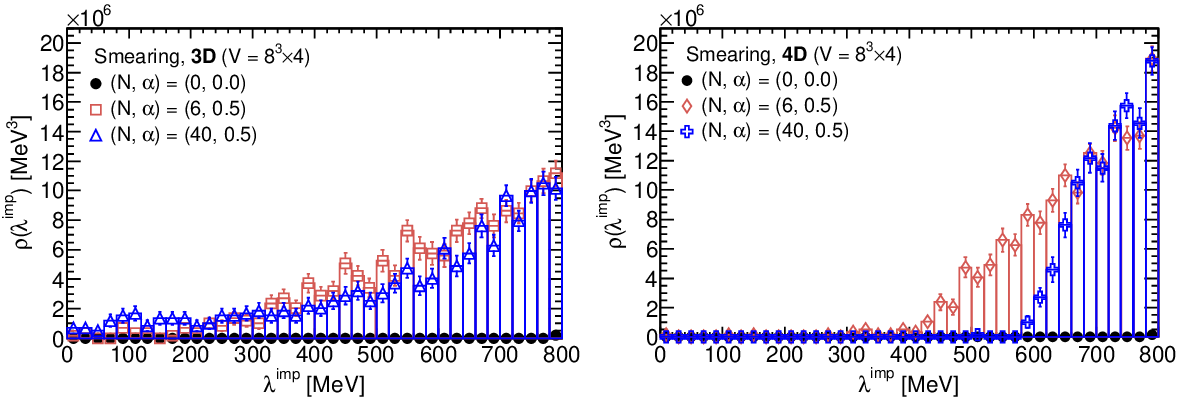}
  \end{center}
  \setlength\abovecaptionskip{-1pt}
  \caption{The spectral density $\rho(\lambda^{\text{imp}})$ at high temperature ($V = 8^{3}\times4$) calculated from the smeared link variables. The results of $(N, \alpha)$ = (0, 0.0) are the same as those shown in Fig.~\ref{fig:Spec_dens_8xx3x4}, in which smearing is not applied.}\label{fig:spect_smear_finite}
\end{figure*}
\begin{figure*}[htbp]
  \begin{center}
   \includegraphics[width=160mm]{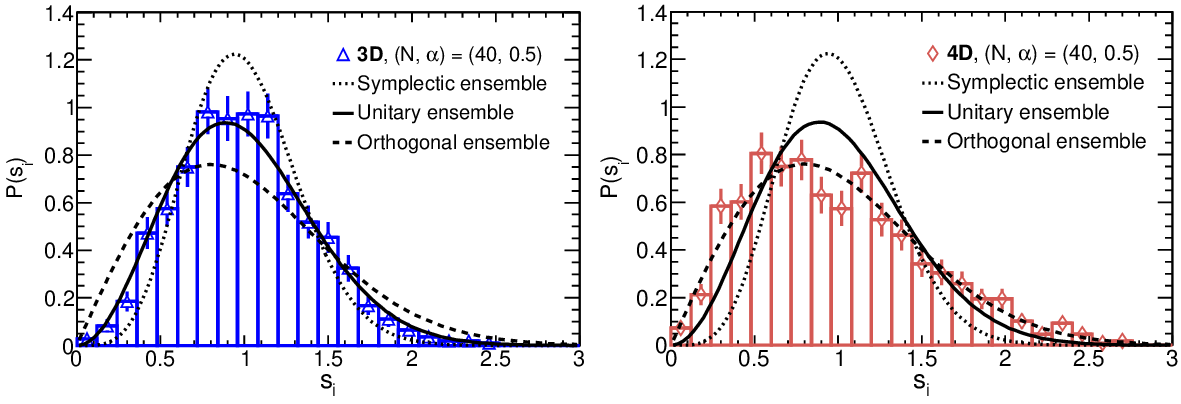}
  \end{center}
  \setlength\abovecaptionskip{-1pt}
  \caption{The distributions P($s_{i}$) of the nearest-neighbor spacing $s_{i}$ at high temperature ($V = 8^{3}\times4$) calculated using the smeared link variables. The parameters for smearing are $(N, \alpha) = (40, 0.5)$.}\label{fig:near_8xx3x4_sm_n40}
\end{figure*}

\clearpage

\section*{Acknowledgments}

The author acknowledges helpful discussions with C. Bonati, M. D'Elia, and F. Negro and their assistance. The author received financial support from the University of Pisa, the Istituto Nazionale di Fisica Nucleare at the University of Pisa, and the Bogoliubov Laboratory of Theoretical Physics at the Joint Institute for Nuclear Research. The author uses the SX-series supercomputer and PC clusters provided by the Research Center for Nuclear Physics (RCNP) and Cybermedia Center at Osaka University alongside the storage element of the Japan Lattice Data Grid at the RCNP. The author extends his appreciation for the computer resources provided for this research.

\section*{Data Availability Statement}

The analyzed data for this study are included in this article. The data tables~\cite{Hasegawa7} supporting this article are accessible on figshare with the digital object identifier: ``\url{https://doi.org/10.6084/m9.figshare.21679526}''. The author recommends that anyone using data and content from this article and reference~\cite{Hasegawa7} for machine learning should give proper citations or comments.

\bibliographystyle{unsrt}
\bibliography{Monopoles_instantons_eta_M.Hasegawa_revised_ver_12Jul2023.bib}

\end{document}